\documentclass[graybox, nosecnum]{svmult}

\usepackage{mathptmx}       
\usepackage{helvet}         
\usepackage{courier}        
\usepackage{type1cm}        
\usepackage{makeidx}         
\usepackage{graphicx}        
\usepackage{multicol}        
\usepackage[bottom]{footmisc}
\usepackage{hyperref}        
\usepackage{soul}            
\usepackage{ulem}

\newcommand{\beqa}{\begin{eqnarray}}
\newcommand{\eeqa}{\end{eqnarray}}

\makeindex             

\begin{document}
\title*{Lattice QCD and Baryon-Baryon Interactions}
\author{Sinya Aoki\thanks{corresponding author} and Takumi Doi}
\institute{Sinya Aoki \at Center for Gravitational Physics and Quantum Information, Yukawa Institute for Theoretical Physics, Kyoto University,  Kyoto 606-8502, Japan, \email{saoki@yukawa.kyoto-u.ac.jp}
\and Takumi Doi \at Interdisciplinary Theoretical and Mathematical Sciences Program (iTHEMS), RIKEN, Wako  351-0198, Japan, \email{doi@ribf.riken.jp}}
%
%
\maketitle
\abstract{
In this chapter, the current status on baryon-baryon interactions such as nuclear forces in lattice Quantum ChromoDynamics (QCD) is reviewed. 
In studies of baryon-baryon interactions in lattice QCD, the most reliable method so far is the potential method, proposed by the
Hadrons to Atomic nuclei from Lattice QCD (HAL QCD) collaboration,  whose formulation, properties and  extensions are  explained in detail.
Using the HAL QCD potential method, potentials between nucleons (proton and neutron, denoted by $N$)
in the derivative expansion have been extracted in various cases.
The lattice QCD results shown in this chapter include
a Leading Order (LO) central potential in the parity-even $NN(^1S_0)$ channel,  LO central and tensor potentials in the parity-even  $NN(^3S_1$-$^3D_1)$ channel, and  a Next-to-Leading Order (NLO) spin-orbit potential as well as LO potentials in the parity-odd channels.
Preliminary results at the almost physical pion and kaon masses,
in addition to exploratory studies on three-nucleon potentials,  are presented.
Interactions between generic baryons including hyperons, made of one or more strange quarks as well as up and down quarks, 
have also been investigated.
Universal properties of   potentials between baryons become manifest in the flavor SU(3) symmetric limit, where masses of three quarks, up, down and strange, are all equal. In particular, it is observed that one bound state, traditionally called the $H$-dibaryon,  appears  in the flavor singlet representation of SU(3).
A fate of the $H$ dibaryon is also discussed with flavor SU(3) breaking taken into account
at the almost physical point.
Finally,   various kinds of dibaryons, bound or resonate states of two baryons, including charmed dibaryons, have been
predicted by lattice QCD simulations
at the almost physical point.}

\section{\textit{1. Introduction}}
\label{sec:intro}
Properties of nuclei and hyper-nuclei are ultimately controlled by  Quantum ChromoDynamics (QCD), which governs the dynamics of quarks and gluons.
It is, however, too challenging to investigate nuclei directly from QCD,
and thus the most important theoretical task is to determine hadron interactions
such as nuclear forces and hyperon forces
without introducing any model parameters other than the gauge coupling constant of QCD and quark masses.
Currently,
the only method which can systematically incorporate non-perturbative nature of QCD dynamics
is the first-principles lattice QCD numerical simulations.
Indeed, not only hadron masses themselves~\cite{Durr:2008zz} but also hadron mass splittings such as a proton-neutron mass difference can be precisely determined in latest lattice QCD calculations~\cite{Borsanyi:2014jba}. 

In lattice QCD, hadron-hadron interactions have been mainly investigated by two methods, the finite volume method~\cite{Luscher:1990ux} and the HAL QCD potential method~\cite{Ishii:2006ec,Aoki:2009ji}. 
In the finite volume method, two-hadron spectra in finite volume are converted to scattering phase shifts through the L\"uscher''s finite volume formula,
while Nambu-Bethe-Salpeter (NBS) wave functions for two hadrons are converted to an interaction kernel or a non-local potential between two hadrons
in the HAL QCD method, with which  scattering phase shifts are extracted through the Schr\"odinger equation.  
Both methods are theoretically equivalent, but they have their own pros and cons in practice.
The finite volume method has been mainly used for meson-meson systems in both center of mass (CM) and moving frames~\cite{Rummukainen:1995vs}.
(See \cite{Briceno:2017max} for a review.)
On the other hand, the HAL QCD potential method has  been applied mainly to  baryon-baryon systems in the CM frame, and an extension to the moving frame has just begun~\cite{Aoki:2021tcu,Akahoshi:2023eku}.
Some reviews are given in~\cite{Aoki:2008hh,Aoki:2010ry,Aoki:2011ep,Aoki:2012tk,Aoki:2013tba,Aoki:2020bew}.

In the past, there were some controversies over the consistency
between the finite volume method and the HAL QCD potential method.
In the case  of nucleon-nucleon ($NN$) interactions
at heavier pion masses, while finite volume spectra predict that both deuteron and dineutron are bound~\cite{Yamazaki:2011nd,NPLQCD:2011naw,NPLQCD:2012mex,Yamazaki:2012hi,Yamazaki:2015asa,Berkowitz:2015eaa,Orginos:2015aya,Wagman:2017tmp},
$NN$ potential in the HAL QCD method indicate that two nucleons are unbound in both channels~\cite{Ishii:2006ec,Aoki:2009ji,Inoue:2011ai,Ishii:2012ssm}.
Then it has been pointed out that incorrect extractions of $NN$ finite volume spectra are responsible for the $NN$ controversy~\cite{Iritani:2016jie,Aoki:2016dmo,Iritani:2017rlk,Aoki:2017byw,HALQCD:2018gyl,Iritani:2018vfn} (See also~\cite{Aoki:2020bew} for a summary),
and this claim is being confirmed by recent studies on $NN$ finite volume spectra~\cite{Francis:2018qch,Horz:2020zvv,Green:2021sxb,Horz:2022glt,Nicholson:2021zwi,Wagman:2021spu} using more sophisticated methods such as the variational method~\cite{Luscher:1990ck}.
Now it is the Lattice QCD community's consensus that $NN$ systems are unbound at heavier pion masses in both channels.   
With the reliability of the HAL QCD method well established,
results on baryon-baryon ($BB$) interactions in the HAL QCD method will be exclusively reported in this chapter.

\section{\textit{2. HAL QCD potential method}} 
In this section, the HAL QCD potential method~\cite{Ishii:2006ec,Aoki:2009ji,Aoki:2012tk} is briefly reviewed.

\subsection{2-1. Basic formulation}
A key quantity in the HAL QCD method is the equal time NBS wave function, defined by
\begin{eqnarray}
\psi_{W}^{B_1B_2}({\bf r}) e^{-W t} \equiv {1\over \sqrt{Z_{B_1} Z_{B_2}}} \sum_{\bf x} \langle \Omega \vert
B_1({\bf x}+{\bf r},t) B_2({\bf x}, t) \vert B_1B_2; W\rangle,
\end{eqnarray}
where
$\langle \Omega \vert$ is the QCD vacuum (bra-)state and
$\vert B_1B_2; W\rangle$ is a QCD eigenstate for two baryons with equal masses $m_{B_1}=m_{B_2}=m_B$,
the CM energy $W= 2\sqrt{{\bf p}_W^2 + m_{B}^2}$ and a relative momentum ${\bf p}_W$.
For simplicity, other quantum numbers such as spin and helicity are suppressed.
A baryon operator $B_{1,2}({\bf x},t)$ is  made of quarks,  which is given by,
e.g., for a local nucleon operator,
\[
B_{1,2}(x)=N_q(x) = \epsilon^{abc} \left( u^{aT}(x) C\gamma_5 d^b(x) \right) q^c(x),
\quad x = ({\bf x},t),
\]
where $C=\gamma_2\gamma_4$ is a charge conjugation matrix, $q=u(d)$ for a proton (neutron). 
Extensions of formula in this section to unequal masses $m_{B_1}\not=m_{B_2}$ are straightforward but some formulae become more complicated. 

If the total energy lies below the inelastic threshold of one meson production as $W < W_{\rm th}:=2 m_B + m_M$, where $m_M$ is a mass of the lightest meson $M$
coupled to this channel, 
the above NBS wave function satisfies the Helmholtz equation  at large $r=\vert {\bf r}\vert>R$ as
\begin{eqnarray}
\left[ p_W^2 +\nabla^2\right] \psi_{W}^{B_1B_2}({\bf r}) \simeq 0, \quad  p_W =\vert {\bf p}_W\vert,
\end{eqnarray}
where $R$  is a scale of the two-baryon interaction, which is short-ranged  in QCD. More precisely,  the NBS wave function for given orbital angular momentum $\ell$ and total spin $s$ behaves asymptotically as
\begin{eqnarray}
 \psi_{W}^{B_1B_2}(r;\ell s)&\simeq& {\sin ( p_W r -\ell\pi/2 +\delta_{\ell s}(p_W) ) \over p_W r} e^{i  \delta_{\ell s}(p_W) }
\end{eqnarray}
for large $r > R$, where $\delta_{\ell s}(p_W)$ is a phase of the QCD S-matrix for $B_1 B_2$ scattering below the inelastic threshold,
which is real thanks to the unitarity of the S-matrix~\cite{Lin:2001ek,CP-PACS:2005gzm}.

The HAL QCD method introduces a non-local but energy-independent potential $U({\bf r}, {\bf r}')$ from the NBS wave function as
\begin{eqnarray}
(E_W - H_0) \psi_{W}^{B_1B_2}({\bf r}) &=& \int d^3r'\, U({\bf r}, {\bf r}')  \psi_{W}^{B_1B_2}({\bf r}'),
\quad E_W  = {p_W^2\over 2\mu}, \ H_0 = -{\nabla^2\over 2\mu}, 
\label{eq:nonlocal U}
\end{eqnarray}
where $\mu = m_B/2$ is a reduced mass, and $W < W_{\rm th}$ is always required.
It is important to note that non-relativistic approximation is not required to define $U({\bf r}, {\bf r}')$ even though eq.~(\ref{eq:nonlocal U})
has a form of the Schr\"odinger equation.

NBS wave functions for $^\forall W < W_{\rm th}$ are necessary to determine the non-local potential $U({\bf r}, {\bf r}')$, which correctly reproduces scattering phase shifts $\delta_{\ell s}(p_W)$ for all $\ell$, $s$ and $p_W$ with $W < W_{\rm th}$.
In lattice QCD simulations, performed in finite boxes, however, only a limited number of NBS wave functions  are available, so that
a direct determination of the non-local potential  is impractical. Instead, one needs to expand the non-local potential in terms of the velocity (derivative) with local coefficient functions as $U({\bf r}, {\bf r}') = V({\bf r},{\bf \nabla}) \delta^{(3)}({\bf r} -{\bf r}')$, where
\begin{eqnarray}
 V({\bf r},{\bf \nabla}) &=&\underbrace{V_0(r) + V_\sigma(r) {\bf \sigma}_1\cdot  {\bf \sigma}_2 + V_{\rm T}(r) S_{12}}_{\rm LO}+\underbrace{V_{\rm LS}(r){\bf L}\cdot{\bf S}}_{\rm NLO} + O(\nabla^2)
\end{eqnarray}
at lowest few orders for the $NN$ case with a given isospin.
Here $V_0(r)$ is a central potential, $V_\sigma$ is a spin dependent central potential with a Pauli matrix $\sigma_i$ acting on a spinor index of an $i$-th baryon, $V_{\rm T}(r)$ is a tensor potential with a tensor operator $S_{12}= 3({\bf r} \cdot{\bf \sigma}_1) ({\bf r} \cdot{\bf \sigma}_2)/r^2-  {\bf \sigma}_1\cdot  {\bf \sigma}_2$,  and $V_{\rm LS}(r)$ is a spin-orbit potential with an angular momentum ${\bf L} ={\bf r}\times {\bf p}$ and a total spin
${\bf S}=({\bf \sigma}_1 +{\bf \sigma}_2)/2$.
Each coefficient function is further decomposed into its flavor components. For example, in the case of $NN$ interactions, the decomposition reads
\begin{eqnarray}
V_X(r) &=& V_X^0(r) + V^\tau_X(r) {\bf \tau}_1\cdot{\bf \tau}_2, \quad X=0,\sigma, {\rm T}, {\rm LS},\cdots,
\end{eqnarray}
where ${\bf \tau}_i$ is the Pauli matrix acting on the SU(2) flavor index of the $i$-th baryon.
The above form of the velocity expansion has already been found in~\cite{Okubo1958} using a similar argument based on symmetries,
though an expansion of the non-local potential is not unique.

The leading order potential is simply obtained from one NBS wave function as
\begin{eqnarray}
V^{\rm LO}({\bf r}) &=& { ( E_W -H_0)  \psi_{W}^{B_1B_2}({\bf r}) \over  \psi_{W}^{B_1B_2}({\bf r})} = V_0(r) +  V_\sigma(r) {\bf \sigma}_1\cdot  {\bf \sigma}_2 + V_{\rm T}(r) S_{12}.
\end{eqnarray}
Solving the Schr\"odinger equation with $V^{\rm LO}({\bf r})$ in an infinite volume, one obtains scattering phase shifts $\delta_{\ell s}^{\rm LO}(q)$,
which satisfy  $\delta_{\ell s}^{\rm LO}(p_W) = \delta_{\ell s}(p_W)$ since $ \psi_{W}^{B_1B_2}({\bf r})$ is a solution to the  Schr\"odinger equation with $V^{\rm LO}({\bf r})$ at $q=p_W$, but $\delta_{\ell s}^{\rm LO}(q) \neq \delta_{\ell s}(q)$ for general $q\not=p_W$.
Hereafter, a momentum $p_W$, which gives a correct phase shift $\delta_{\ell s}(p_W)$, is called an ``anchor".
If NBS wave functions at two different energies $W_1$ and $W_2$ are  obtained, the non-local potential can be determined at the NLO in the velocity expansion. The NLO scattering phase shifts $\delta^{\rm NLO}_{\ell s}(q)$ not only satisfy $\delta^{\rm NLO}_{\ell s}(p_{W_1})=\delta_{\ell s}(p_{W_1})$ and $\delta^{\rm NLO}_{\ell s}(p_{W_2})=\delta_{\ell s}(p_{W_2})$, but also give good approximations of $\delta_{\ell s}(q)$ between two anchors,
$p_{W_1}$ and $p_{W_2}$. 
In this way, predictions of scattering phase shifts in the HAL QCD method can be improved  by increasing order of the velocity expansion of the non-local potential. See some explicit demonstrations in Refs.~\cite{Aoki:2019pnq,Aoki:2021ahj}.

Since NBS wave functions cannot be obtained directly in lattice QCD,
a correlation function for two baryons is considered instead,
\begin{eqnarray}
F_{J}^{B_1B_2}({\bf r}, t) = \sum_{\bf x} \langle \Omega\vert B_1({\bf x}+{\bf r},t+t_0) B_2({\bf x},t + t_0) {J}^\dagger_{B_1B_2}(t_0)\vert \Omega\rangle,
\end{eqnarray}
 where ${J}^\dagger_{B_1B_2}(t_0)$ is a source operator which creates two baryon states at time $t_0$.
 If a time separation $t$ is large enough to suppress inelastic contributions to the correlation function, it is shown that
 \begin{eqnarray}
F_{J}^{B_1B_2}({\bf r}, t) &=& \sum_n A_{J,n}^{B_1B_2}  \psi_{W_n}^{B_1B_2}({\bf r}) e^{-W_n t} +\cdots, 
\end{eqnarray}
where $A_{J,n}^{B_1B_2} =\sqrt{Z_{B_1} Z_{B_2}}\  \langle B_1B_2;W_n \vert {J}_{B_1B_2}^\dagger (0)\vert \Omega\rangle$ and ellipses represent contributions from inelastic states such as $B_1 B_2 M$, which are suppressed at least as $e^{- W_{\rm th} t}$.
In the limit of $t \to\infty$, the correlation function reduces to the NBS wave function with the lowest energy as
\begin{eqnarray}
\lim_{t \to\infty} F_{J}^{B_1B_2}({\bf r}, t) &=& A_{J,0}^{B_1B_2}   \psi_{W_0}^{B_1B_2}({\bf r}) e^{-W_0 t}
+ O\left( e^{-W_{n\not=0} t} \right),
\label{eq:groundF}
\end{eqnarray}
where $W_0$ is the lowest energy of $B_1B_2$ states.
$W_{n\not=0} - W_0 = O( L^{-2} )$ in lattice QCD on a finite box with a volume $L^3$,
where we neglect the effect of a binding energy (if there exists a bound state) for simplicity.
(See Ref.~\cite{Gongyo:2018gou} if a bound state appears.) 
The leading order potential is obtained as
\begin{eqnarray}
V^{\rm LO}({\bf r}) &=& \lim_{t\to\infty} V^{\rm LO}({\bf r},t), \quad
V^{\rm LO}({\bf r}, t) :=
{ ( E_W -H_0)  F_J^{B_1B_2}({\bf r},t) \over  F_{J}^{B_1B_2}({\bf r}, t)} ,
\label{eq:pot_naive}
\end{eqnarray}
which is independent on the source operator$J^\dagger_{B_1B_2}$, since a source dependent constant $A_{J,0}^{B_1B_2}$ is cancelled in the above ratio as $t\to\infty$.
For the above procedure to work, the ground state saturation in $F_J^{B_1B_2}$, eq.~(\ref{eq:groundF}), must be satisfied by taking a large $t$.
In practice, however, $F_J^{B_1B_2}$ becomes very noisy at large enough $t$, in particular for two baryons.
To overcome this difficulty, an alternative extraction of potentials has been introduced.

A ratio of correlation functions is defined by
\begin{eqnarray}
R_J^{B_1B_2}({\bf r},t) := {F_J^{B_1B_2}({\bf r},t) \over G_{B_1}(t) G_{B_2}(t)}, \quad
G_B(t) :=\sum_{\bf x} \langle \Omega\vert B({\bf x},t) \bar B({\bf 0},0)\vert \Omega \rangle \simeq Z_B e^{-m_B t} + \cdots, ~~~~~
\end{eqnarray}
which behaves
\begin{eqnarray}
R_J^{B_1B_2}({\bf r},t) &\simeq&\sum_n {A_{J,n}^{B_1B_2}\over Z_{B_1} Z_{B_2}}  \psi_{W_n}^{B_1B_2}({\bf r}) e^{-\Delta W_n t},
\quad \Delta W_n := W_n - 2 m_B,
\end{eqnarray}
for $ t \gg 1/W_{\rm th}$, where inelastic contributions can be neglected. Using
\begin{eqnarray}
\Delta W ={p_{W}^2\over m_B} -{(\Delta W)^2\over 4 m_B}, \quad 
(E_W - H_0)  \psi_{W}^{B_1B_2}({\bf r}) = V({\bf r},{\bf \nabla} )  \psi_{W}^{B_1B_2}({\bf r}), 
\end{eqnarray}
the ration is shown to satisfy 
\begin{eqnarray}
\left\{ - H_0 -{\partial \over \partial t} +{1\over 4 m_B} {\partial^2\over \partial t^2} \right\}
R_J^{B_1B_2}({\bf r},t) =  V({\bf r},{\bf \nabla} )  R_J^{B_1B_2}({\bf r},t) ,
\label{eq:t-dep}
\end{eqnarray}
where $ V({\bf r},{\bf \nabla} ) $ can be extracted in terms of the velocity expansion as mentioned before.
The required condition, 
$t \gg 1/W_{\rm th}$ (saturation of elastic states),
is much easier to be achieved than the condition in the usual finite volume method,
$ t \gg 1/(W_1-W_0) \simeq O(L^2)$ (saturation of the ground state).
Eq.~(\ref{eq:t-dep}) is called the time dependent HAL QCD method~\cite{Ishii:2012ssm}.

\subsection{2-2. An extension: coupled channel potentials}
The HAL QCD potential method can be extended to certain types of coupled channel systems~\cite{Aoki:2011gt,Sasaki:2010bi}.
For simplicity, a case that $X_1+X_2\to Y_1+Y_2$ scattering occurs in addition to an elastic scattering $X_1+X_2\to X_1+X_2$ is considered, where $X_{1,2},Y_{1,2}$ are one-particle states of  hadrons, and $  m_{X_1}+m_{X_2}  <  m_{Y_1} + m_{Y_2} < W < W_{\rm th}$ with $W$ and $W_{\rm th}$ being the total energy and the inelastic threshold of the coupled channel system, respectively.  For simplicity, $ m_{X_1}=m_{X_2}=m_X$ and $m_{Y_1}=m_{Y_2}=m_Y$ are assumed in the following.

In this case, NBS wave functions are generalized as
\begin{eqnarray}
\Psi^{X}_W ({\bf r} )e^{-W t}  &=& \frac{1}{\sqrt{Z_{X_1} Z_{X_2}}} \sum_{\bf x} \langle \Omega \vert X_1({\bf x} +{\bf r},t) X_2 ({\bf x},t)\vert X+Y; W \rangle ,  \\
\Psi^{Y}_W ({\bf r} )e^{-W t}  &=& \frac{1}{\sqrt{Z_{Y_1} Z_{Y_2}}} \sum_{\bf x} \langle \Omega \vert Y_1({\bf x} +{\bf r},t) Y_2 ({\bf x},t)\vert X+Y; W \rangle ,
\end{eqnarray}
where $\vert X+Y; W\rangle$ is a QCD eigenstate in the coupled channel system, which may be expressed as
\begin{eqnarray}
\vert X+Y;W\rangle &=& c_{X} \vert X_1X_2;W\rangle + c_Y  \vert Y_1Y_2;W\rangle +\cdots, \\
\vert H_1H_2;W\rangle &=&  \sum_{\vert{\bf p}\vert = p_W^H} \vert H_1, {\bf p} \rangle_{\rm in} \otimes   \vert H_2, -{\bf p} \rangle_{\rm in}
\end{eqnarray}
for $H=X,Y$. 
Here $\vert H_{1,2}, {\bf p} \rangle_{\rm in}$ is an in-state for a hadron $H_{1,2}$ with a momentum ${\bf p}$, and  
$p_W^H$ is a magnitude of a relative momentum in the channel $H$, given by $W=2\sqrt{(p_W^H)^2+m_H^2}$ for $H=X,Y$.

A coupled channel non-local potential is defined  by
\begin{eqnarray}
\left[ {(p_W^H)^2\over 2\mu_H} +{\nabla^2\over 2\mu_H} \right] \Psi_W^H({\bf r}) = \sum_{H'=X,Y}\int d^3r'\, U^H{}_{H'}({\bf r},{\bf r'} )  \Psi_W^{H'}({\bf r'}), 
\end{eqnarray}
where $\mu_H= m_H/2$ is  a reduced mass in a channel $H$.
At the leading order of the velocity expansion,
two pairs of NBS wave functions at two different energies $W_1$ and $W_2$ satisfy
\begin{eqnarray}
K^H{}_{W_i}({\bf r}) &:=& \left[ {(p_{W_i}^H)^2\over 2\mu_H} +{\nabla^2\over 2\mu_H} \right] \Psi_{W_i}^H({\bf r}) 
= \sum_{H'=X,Y} V^H{}_{H'}({\bf r} )  \Psi_{W_i}^{H'}({\bf r}), \quad i=1,2,
\end{eqnarray}
where $ V^H{}_{H'}({\bf r} )$ is a local potential matrix at the leading order, which is solved as
\begin{eqnarray}
\left(
\begin{array}{cc}
 V^X{}_X({\bf r}) &  V^X{}_Y({\bf r})    \\
 V^Y{}_X({\bf r})  & V^Y{}_Y({\bf r})    \\
\end{array}
\right)
&=&
\left(
\begin{array}{cc}
 K^X{}_{W_1}({\bf r}) &  K^X{}_{W_2}({\bf r})    \\
 K^Y{}_{W_1}({\bf r})  & K^Y{}_{W_2}({\bf r})    \\
\end{array}
\right)
\left(
\begin{array}{cc}
 \Psi^X_{W_1}({\bf r}) &  \Psi^X_{W_2}({\bf r})    \\
 \Psi^Y_{W_1}({\bf r})  & \Psi^Y_{W_2}({\bf r})    \\
\end{array}
\right)^{-1} .
\end{eqnarray}
Once the coupled channel potential matrix $V^H{}_{H'}({\bf r})$ is obtained, physical observables such as scattering phase shifts and inelasticities 
can be extracted as a function of $W$ by solving a coupled channel Schr\"odinger equation in an infinite volume  with appropriate boundary conditions.

As in the case of the single channel potential, NBS wave functions are extracted in lattice QCD from correlation functions as
\begin{eqnarray}
F_J^H({\bf r}, t) &:=& \sum_{\bf x} \langle \Omega\vert H_1({\bf x}+{\bf r},t) H_2({\bf x},t) J^\dagger_H (0) \vert \Omega\rangle \nonumber \\
&\simeq& A_{J,0}^H \Psi_{W_0}^H({\bf r})e^{-W_0 t} + A_{J,1}^H \Psi_{W_1}^H({\bf r})e^{-W_1 t} + O(e^{- W_2 t})
\end{eqnarray}
for large $t$, where $W_0 < W_1 < W_2$ are 3 lowest eigen-energies.
Using two different source operators, $J_{1,H}^\dagger$ and $J_{2,H}^\dagger$, we may extract 
$ \Psi_{W_0}^H({\bf r})$ and $ \Psi_{W_1}^H({\bf r})$ for $H=X,Y$, but this may be a harder task than an extraction of  
$ \Psi_{W_0}^X({\bf r})$ in the single channel.

Therefore the time dependent method is also employed in the coupled channel case. 
Using
\begin{eqnarray}
R_{J_i}^H({\bf r}, t) &:=& {F_{J_i}^H({\bf r},t)\over G_{H_1}(t) G_{H_2}(t)}
\simeq \sum_n  {A_{J_i,n}^H\over Z_{H_1} Z_{H_2}} \Psi_{W_n}^H({\bf r}) e^{- \Delta W_n^H t}, 
\end{eqnarray}
for $H=X,Y$ and $ i=1,2$,
where $J_i$ represent two different source operators, $\Delta W_n^H=W_n - m_{H_1}-m_{H_2}$, and 
$t$ is taken to be large enough to ignore inelastic contributions,  it is easy to show
\begin{eqnarray}
\left( {\nabla^2\over 2\mu_H}-{\partial\over \partial t} + {1\over 8\mu_H} {\partial^2\over \partial t^2} \right) R_{J_i}^H({\bf r}, t) 
&\simeq & \sum_{H'} \Delta^{H}{}_{H'}(t) \int d^3 r' \ U^H{}_{H'}({\bf r}, {\bf r'}) R_{J_i}^{H'}({\bf r'}, t),~~~~
\label{eq:t-dep_C}
\end{eqnarray}
where
\begin{eqnarray}
 \Delta^{H}{}_{H'} (t) = \sqrt{ Z_{H'_1} Z_{H'_2}\over Z_{H_1} Z_{H_2}} { e^{ - (m_{H'_1} + m_{H'_2}) t}\over e^{ - (m_{H_1} + m_{H_2}) t}},
\end{eqnarray}
which is necessary to correct differences in masses and $Z$ factors between two channels.
Denoting the left-hand side of eq.~(\ref{eq:t-dep_C}) as $K^H{}_{J_i}({\bf r}, t)$, the LO potential, $U^H{}_{H'}({\bf r}, {\bf r'}) = V^H{}_{H'}({\bf r})
\delta^{(3)}({\bf r} - {\bf r'})$, is extracted as
\begin{eqnarray}
\left(
\begin{array}{cc}
 V^X{}_X({\bf r}) &  \Delta^X{}_Y(t) V^X{}_Y({\bf r})    \\
 \Delta^Y{}_X(t)  V^Y{}_X({\bf r})  & V^Y{}_Y({\bf r})    \\
\end{array}
\right)
&=&
\left(
\begin{array}{cc}
 K^X{}_{J_1}({\bf r},t) &  K^X{}_{J_2}({\bf r},t)    \\
 K^Y{}_{J_1}({\bf r},t)  & K^Y{}_{J_2}({\bf r},t)    \\
\end{array}
\right)
\left(
\begin{array}{cc}
 R^X_{J_1}({\bf r},t) &  R^X_{J_2}({\bf r},t)    \\
 R^Y_{J_1}({\bf r},t)  & R^Y_{J_2}({\bf r},t)    \\
\end{array}
\right)^{-1} . ~~~~~~~
\end{eqnarray}

\section{\textit{3. NN interactions}}
In this section, results on nucleon-nucleon ($NN$) interactions in the HAL QCD potential method are summarized.   

\subsection{3-1. Central and tensor interactions in parity-even channels}
In the parity-even channels between two nucleons, $^1S_0$ (the isospin triplet and the spin singlet $S$ wave) and $^3S_1 - {}^3D_1$ (the isospin singlet and the spin triplet $S$ and $D$ waves)
have been investigated by the LO analysis.
Since the statistical fluctuations for two nucleons increase as a pion mass decreases,
  the studies have been performed mainly at heavy pion masses.

The LO potential for $^1S_0$ is simply obtained from eq.~(\ref{eq:t-dep}), while the leading order potential in the $^3S_1-{}^3D_1$ channel 
has two terms, the central potential and the tensor potential as
\begin{eqnarray}
\left( - H_0 + {\partial \over \partial t} +{1\over 4 m_N} {\partial^2\over \partial t^2} \right) R_J^{NN}({\bf r}, t) \simeq 
\left[ V^I_{\rm C}(r) + V^I_{\rm T}(r) S_{12} \right] R_J^{NN}({\bf r}, t) ,
\end{eqnarray}
where $I=0$ represents a total isospin of two nucleons for the $^3S_1 - {}^3D_1$ channel.
To disentangle a central potential $V^I_{\rm C}(r)$ and a tensor potential $V^I_{\rm T}(r)$, $R_J^{NN}({\bf r}, t) $ is decomposed into two contributions,
$P^{A_1^+} R_J^{NN}({\bf r}, t) $ and $(1- P^{A_1^+}) R_J^{NN}({\bf r}, t)$, where
$P^{A_1^+} $ is a projection operator to the $A_1^+$ representation of the cubic group while $1- P^{A_1^+}$ is a projection orthogonal to $P^{A_1^+} $.   
Since the $A_1^+$ representation of the cubic group on a cubic box contains $\ell=0,4,6$ partial waves, 
$P^{A_1^+} R_J^{NN}({\bf r}, t) $  and $(1- P^{A_1^+}) R_J^{NN}({\bf r}, t)$ are expected to be dominated by the $\ell=0$ component (the $S$ wave) and the $\ell=2$ component (the $D$ wave) at low energies, respectively.
These expectations are indeed confirmed numerically~\cite{Aoki:2009ji}.
Using these two components, $V^I_{\rm C}(r)$ and $V^I_{\rm T}(r)$ are extracted by solving linear equations, as in the couple channel case. 
 For a more sophisticated partial wave decomposition, see~\cite{Miyamoto:2019jjc}.
 
 \begin{figure}[h]
  \centering
  \includegraphics[angle=270,width=0.49\textwidth]{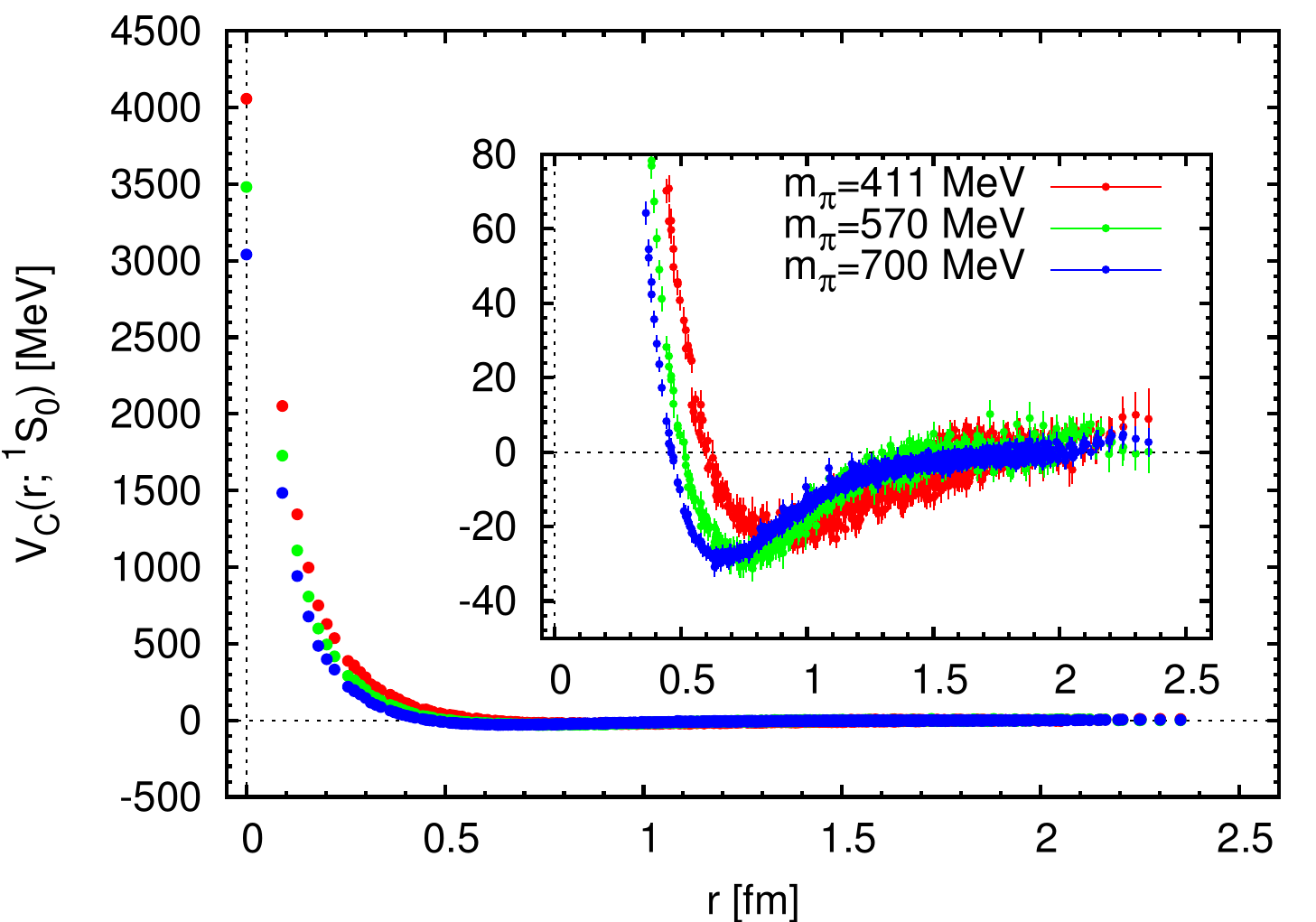}
  \includegraphics[angle=270,width=0.49\textwidth]{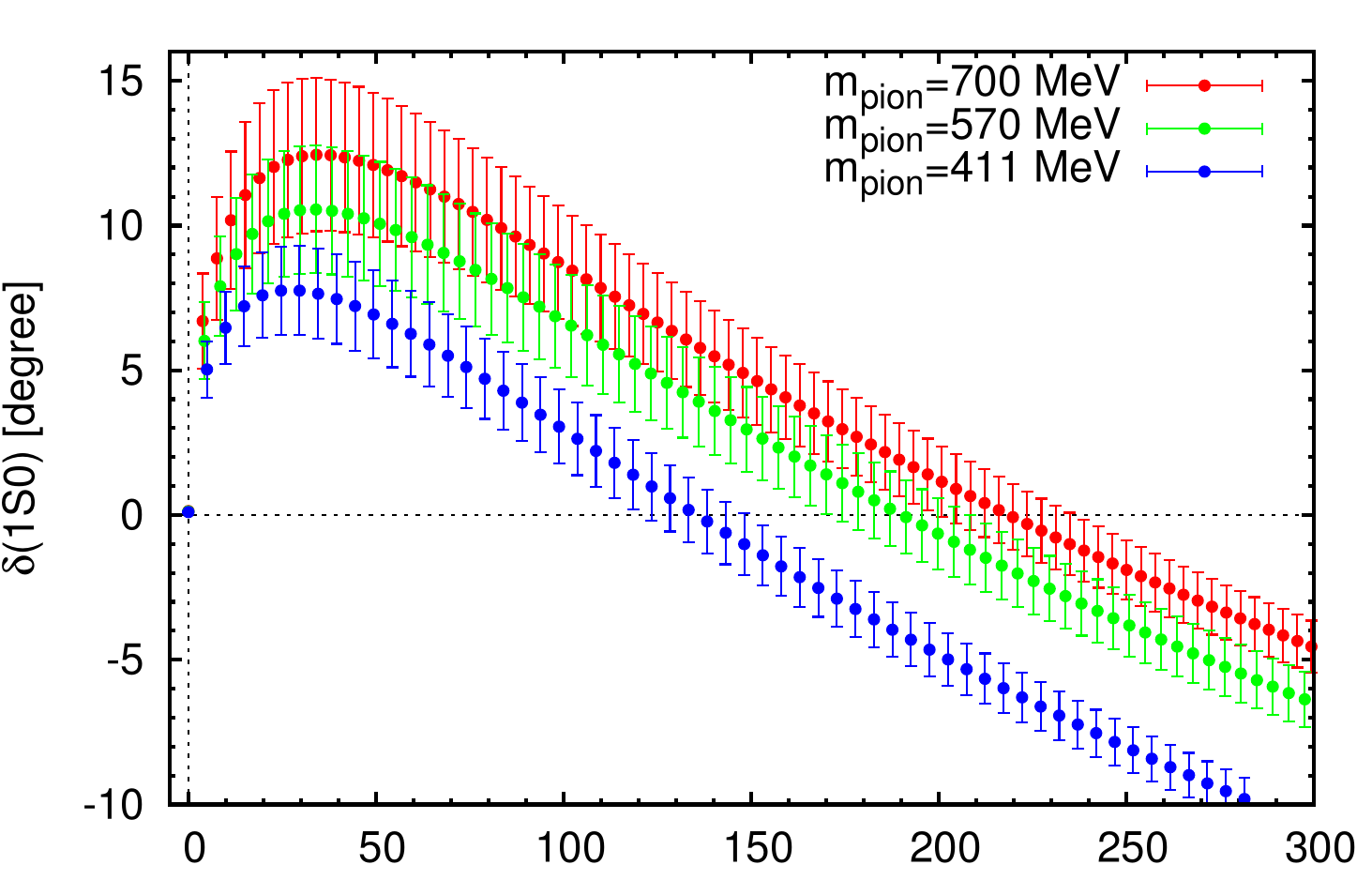}\\
  \caption{(Left) The $NN$ central potential $V^1_{\rm C}(r)$ in the $^1S_0$ channel, obtained from (2+1)-flavor lattice QCD
    at $m_{\pi} \simeq$ 411 (red), 570 (green), 701 (blue) MeV.
    (Right) Corresponding scattering phase shifts as a function of energy. 
     Figures are  taken from~\cite{Ishii:2013ira}.
  }
\label{fig:NN_singlet} 
\end{figure}

\begin{figure}[h]
  \centering
  \includegraphics[angle=270,width=0.49\textwidth]{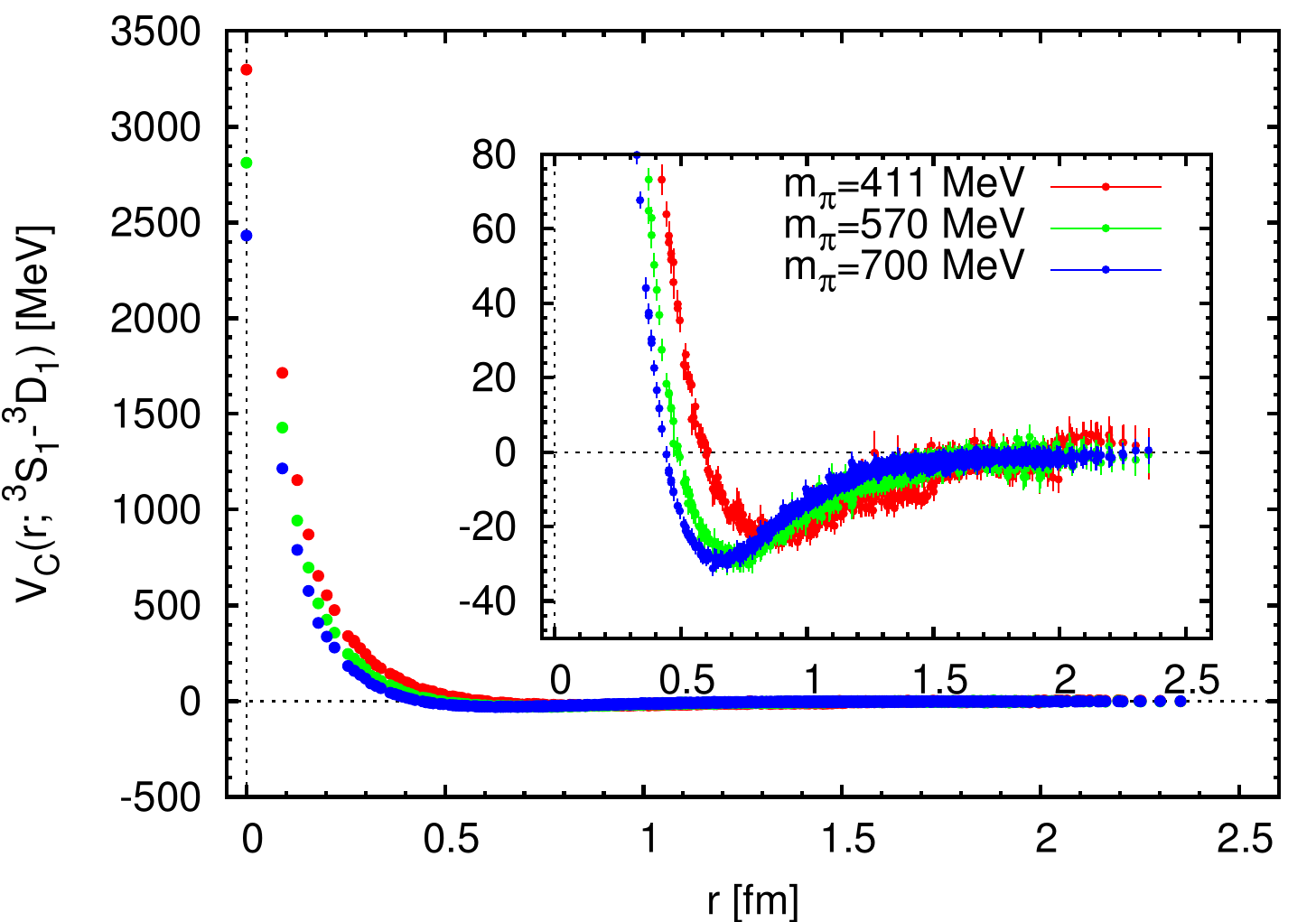}
  \includegraphics[angle=270,width=0.49\textwidth]{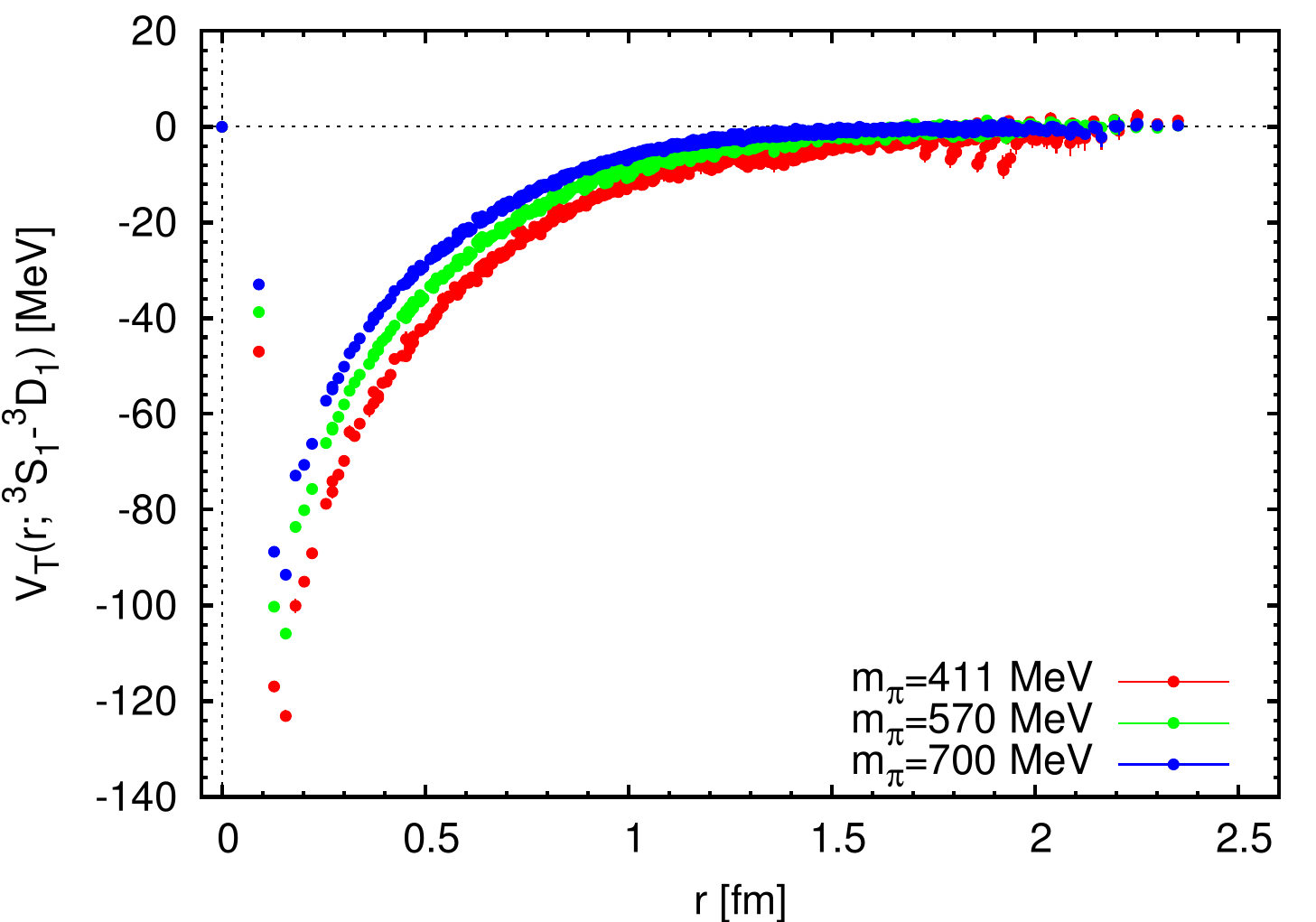}
  \caption{The $NN$ potential in the $^3S_1- ^3D_1$ channel, obtained from (2+1)-flavor lattice QCD
    at $m_{\pi} \simeq$ 411 (red), 570 (green), 701 (blue) MeV.
    (Left) The central potential $V^0_{\rm C}(r)$. (Right) The tensor potential $V^0_{\rm T}(r)$.
     Figures are  taken from~\cite{Ishii:2013ira}.
  }
\label{fig:NN_triplet} 
\end{figure}

 Fig.\ref{fig:NN_singlet} (Left) shows  the central potential in the $^1S_0$ channel, obtained in (2+1)-flavor lattice QCD~\cite{Ishii:2013ira} 
using ensembles generated by the PACS-CS Collaboration at the lattice spacing $a\simeq 0.091$ fm ($a^{-1} = 2.176(31)$ GeV) on $32^3 \times 64$~\cite{PACS-CS:2008bkb}.
Quark masses of these ensembles corresponds to $(m_\pi, m_N) \simeq (701,1584)$ MeV (blue),  $(570,1412)$ MeV (green) and $(411,1215)$ MeV (red).
The central potential at each pion mass reproduces qualitative features of the phenomenological $NN$ potential, namely, 
a repulsive core at short distance surrounded by  an attractive well at medium and long distances.
A range of a tail structure at long distance in the central potential becomes wider as pion mass decreases.
This behavior may be understood from a viewpoint of the one-pion exchange between nucleons.
At short distance, on the other hand, a height of the repulsive core increases as pion mass decreases.
This pion mass dependence of the short range repulsion could be explained by the one-gluon-exchange in the quark model, whose strength is proportional to an inverse of the constituent quark mass.  
Fig.\ref{fig:NN_singlet} (Right) shows the scattering phase shifts, extracted through the Schr\"odinger equation in the infinite volume 
with the corresponding central potential $V^1_{\rm C}(r)$, which indicates that there is no bound state in the channel  at this range of pion masses.
 While behaviors of scattering phase shifts are qualitatively similar to experimental ones, the strength of the attraction is weaker due to heavier pion masses.
Therefore, it is expected that experimental data are reproduced at the physical pion mass up to systematic errors such as the lattice artifact.

Central and tensor potentials, $V^0_{\rm C}(r)$ (Left) and $V^0_{\rm T}(r)$ (Right), in the $^3S_1-^3D_1$ channel from same ensembles
are presented in Fig.\ref{fig:NN_triplet}.
While the central potential has similar shape and quark mass dependence to those in the $^1S_0$  channel,
the tensor potential is strong with a negative sign at all distances, and becomes stronger at lighter pion masses.  

\begin{figure}[h]
  \centering
  \includegraphics[width=0.49\textwidth]{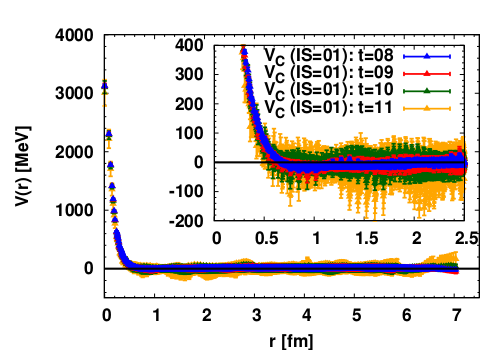}
  \includegraphics[width=0.49\textwidth]{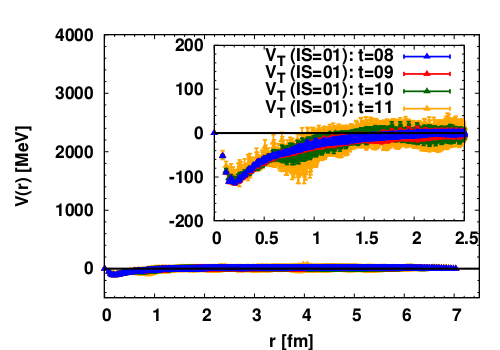}
  \caption{The $NN$ potential in the $^3S_1- ^3D_1$ channel, obtained at $t=8-11$, using 
   the (2+1)-flavor lattice QCD ensemble with $m_{\pi} \simeq 146$ MeV at $a\simeq 0.0846$ fm on a $96^4$ lattice.
    (Left) The central potential $V^0_{\rm C}(r)$. (Right) The tensor potential $V^0_{\rm T}(r)$.
     Figures are  taken from~\cite{Doi:2017zov}.
  }
\label{fig:NN_K} 
\end{figure}
As mentioned before, it is necessary to extract $NN$ potentials at the physical pion mass, in order to confirm whether $NN$ interactions
in the HAL QCD method correctly reproduce experimental results including the binding energy of deuteron, up to systematic errors such as lattice artifacts. As the pion mass decreases, however, statistical fluctuations become larger and larger, so that reliable extractions of $NN$ potentials become difficult.
For example, 
Fig.~\ref{fig:NN_K} shows LO $NN$ potentials in the $^3S_1-^3D_1$ channel, $V^0_{\rm C}(r)$ (Left) and $V^0_{\rm T}(r)$ (Right), obtained at $m_\pi \simeq 146$ MeV (almost physical pion mass)~\cite{Doi:2017zov}, which are much noisier than those 
in Fig.~\ref{fig:NN_triplet} at heavier pion masses.
Note that large fluctuations of these potentials are mainly caused by contaminations from higher partial waves such as $\ell=4$,
as evident form comb-like structures, which could be reduced by the partial wave decomposition method~\cite{Miyamoto:2019jjc}. 

\subsection{3-2. Central, tensor and spin-orbit interactions in parity-odd channels}
In the next-to-leading order (NLO) of the derivative expansion, there appears 
spin-orbital forces (LS) in $NN$ potentials,
which are known to play important roles in the LS-splittings of nuclear spectra and nuclear magic numbers.
It is also argued that the LS force in the $^3P_2-^3F_2$ channel is responsible for the $P$-wave superfluidity in neutron stars,
which affects a cooling process of neutron stars.

LO and NLO potentials in parity-odd channels  have been calculated in
2-flavor lattice QCD with a very heavy quark mass corresponding to  $(m_\pi, m_N)\simeq (1133,2158)$ MeV,    
at $a\simeq 0.156$ fm on a $16^3\times 32$ lattice~\cite{Murano:2013xxa}.
To access parity-odd channels in lattice QCD, source operators should have non-zero momenta, which require more computational costs and
make data noisier. To overcome these difficulties, heavy pion mass was used in this study.

\begin{figure}[h]
  \centering
    \includegraphics[width=0.49\textwidth]{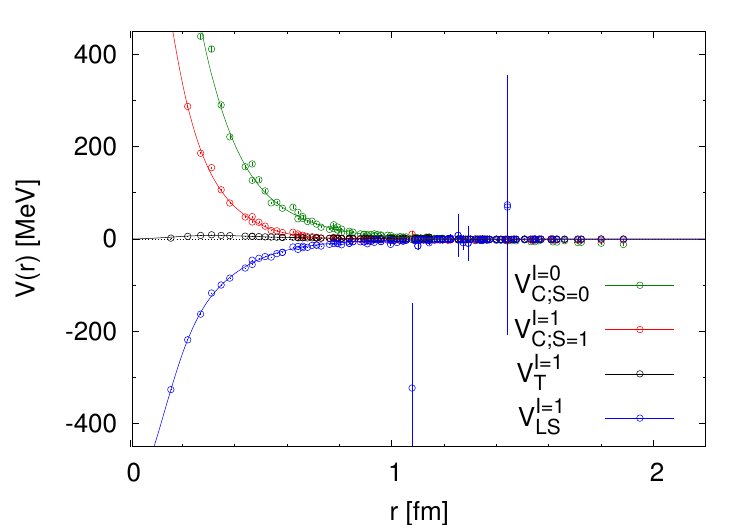}
  \includegraphics[width=0.49\textwidth]{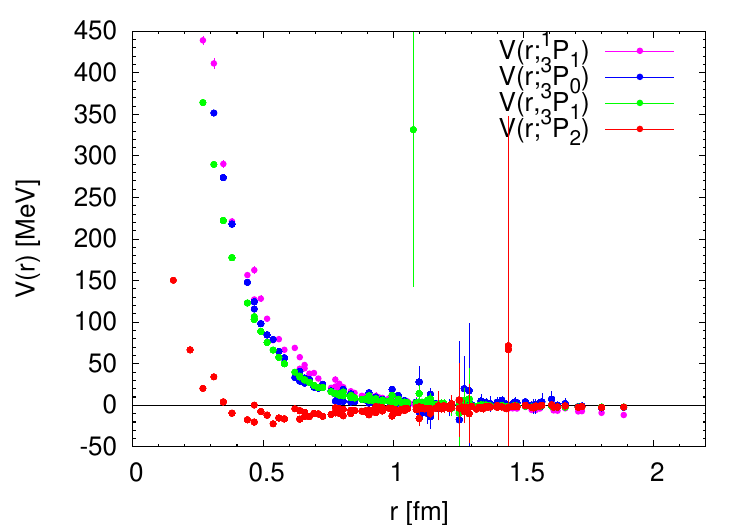}
  \caption{(Left) Central ($S=0$  and $1$), tensor  and spin-orbit potentials
    in parity-odd channels.
    (Right) Potentials  for the $^1P_1$, $^3P_0$,  $^3P_1$ and  $^3P_2$
    channels.
     Figures are  taken from~\cite{Murano:2013xxa}.
  }
\label{fig:pot_odd} 
\end{figure}
Fig.~\ref{fig:pot_odd} (Left) gives LO potentials, $V_{\rm C}^{I=0, S=0}(r)$ (green),   $V_{\rm C}^{I=1, S=1}(r)$ (red) and
$V_{\rm T}^{I=1}(r)$ (black), as well as the LS potential at NLO, $V_{\rm LS}^{I=1} (r)$ (blue).
It is observed that both central potentials  $V_{\rm C}^{I=0, S=0}(r)$ and  $V_{\rm C}^{I=1, S=1}(r)$ are repulsive at all distances and
the tensor potential $V_{\rm T}^{I=1}(r)$ is positive but very weak compared to central potentials,  while the LS potential $V_{\rm LS}^{I=1} (r)$
is strong and negative at all distances. These features are qualitatively consistent with those of phenomenological potentials.
Potentials in $^1P_1, ^3P_0, ^3P_1, ^3P_2$ channels are reconstructed from these 4 potentials as
$V (r; {^1P_1}) =V^{I=0,S=0}_C(r)$, $V (r; {^3P_0}) =V^{I=1,S=1}_C(r) -4 V_{\rm T}^{I=1}(r) - 2 V_{\rm LS}^{I=1}(r)$, 
$V (r; {^3P_1}) =V^{I=1,S=1}_C(r) +2 V_{\rm T}^{I=1}(r) -  V_{\rm LS}^{I=1}(r)$, and
$V (r; {^3P_2}) =V^{I=1,S=1}_C(r) -{2\over 5} V_{\rm T}^{I=1}(r) + V_{\rm LS}^{I=1}(r)$.  
Fig.~\ref{fig:pot_odd} (Right) shows  $V (r; {^1P_1})$ (magenta),  $V (r; {^3P_0})$ (blue), $V (r; {^3P_1})$ (green) and $V (r; {^3P_2})$ (red).

\begin{figure}[h]
  \centering
    \includegraphics[width=0.49\textwidth]{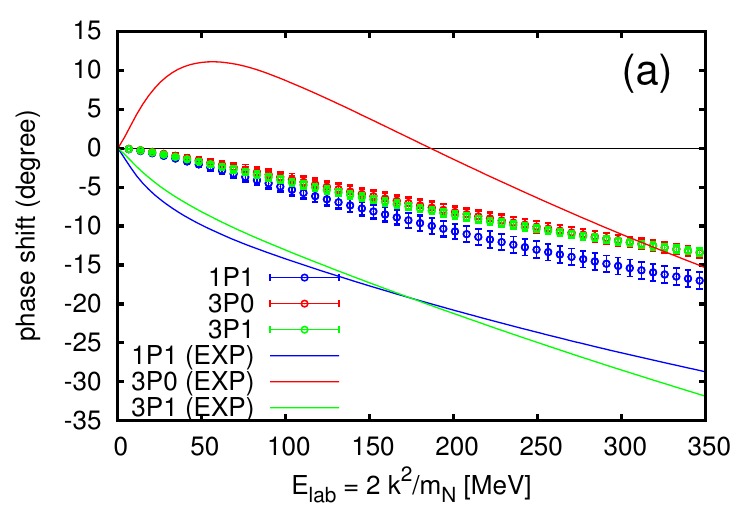}
  \includegraphics[width=0.49\textwidth]{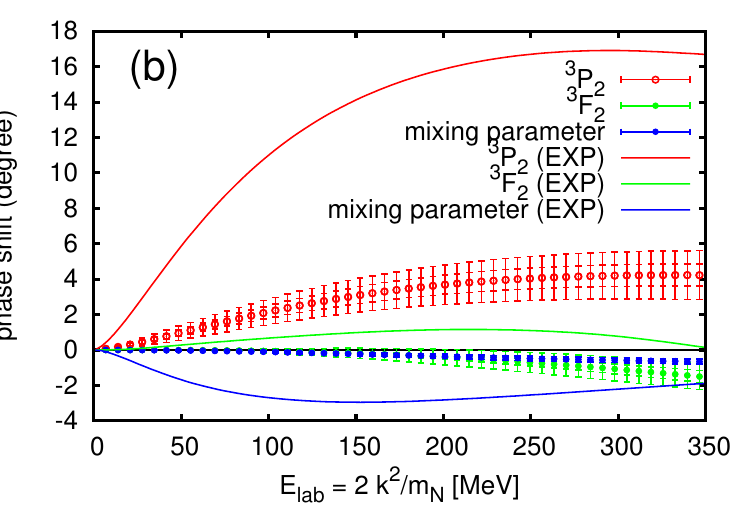}
  \caption{(Left) Scattering phase shifts in the $^1P_1$, $^3P_0$ and $^3P_1$ channels as a function of energy, together with
    experimental data for comparisons.
    (Right) Scattering phase shifts and  mixing parameters (with Stapp's convention)
    in the $^3P_2$--$^3F_2$ channel as a function of energy, together  with experimental data for comparisons.
     Figures are  taken from~\cite{Murano:2013xxa}.
  }
\label{fig:phase_odd} 
\end{figure}
Scattering observables (scattering phase shifts and mixing angles) extracted from these potentials are plotted in Fig.~\ref{fig:phase_odd}.
Compared with experimental phase shifts, qualitative features of  scattering phase shifts are roughly reproduced, while their magnitudes are much smaller
probably due to the much heavier pion mass in this study.
For the scattering phase shift in the $^3P_0$  channel (red  data in the left figure), an attractive behavior at low energies is missing compared with the experimental one (red line), which is likely due to a weak tensor force $V_{\rm T}^{I=1}(r)$ (black in the left of Fig.~\ref{fig:pot_odd}) caused by the much heavier pion mass.  
The most interesting feature is an attractive behavior of the phase shift in the $^3P_2$ channel  (red in the right of Fig.~\ref{fig:phase_odd}),
which originates from the strong attraction of the LS potential $ V_{\rm LS}^{I=1}(r)$ (blue in the left of Fig.~\ref{fig:pot_odd}).
As mentioned, this LS interaction is relevant to the pairing correlation of neutrons and possible $P$-wave superfluidity in neutron stars.

A next step in future studies will be calculations of parity-odd potentials at lighter pion masses, in order to see
the tensor potential and the LS potential become larger in magnitudes so as to reproduce attractive behaviors of scattering phase shifts 
in $^3P_0$ channel and $^3P_2$ channel, respectively. 

\subsection{3-3. Three-nucleon interactions}
Not only two-body $NN$ interactions but also three-body $NNN$ interactions are necessary for nuclear physics.
Three-nucleon interactions play important roles in nuclear spectra and structures such as binding energies of light nuclei and properties of neutron-rich
nuclei. In addition, they are important pieces for properties of nuclear matters such as an equation of state (EoS) at high density, relevant to structures of neutron stars and nucleosynthesis at binary neutron star mergers.
Although there have been many investigations for constructions of three-nucleon interactions  
by phenomenological~\cite{Coon:2001pv,Pieper:2001ap}  or chiral EFT~\cite{Weinberg:1992yk,Epelbaum:2008ga,Machleidt:2011zz,Hammer:2019poc} approaches, it would be most desirable to determine three-nucleon interactions directly from the first principles
lattice QCD.

Unlike $NN$ potentials, which can be derived with relativistic kinematics in the HAL QCD method,
derivations of three-nucleon potentials are currently restricted to non-relativistic kinematics in the HAL QCD method.
See Refs.~\cite{Aoki:2012bb,Aoki:2013cra} for definitions of $n$-body potentials in the HAL QCD method.

A correlation function for three nucleons is defined by
\beqa
F_{J_{3}}^{3N}({\bf r}, {\bf \rho}, t)=\sum_{\bf R}\langle \Omega\vert N({\bf x}_1,t+t_0)  N({\bf x}_2,t+t_0) N({\bf x}_3, t+t_0) J_{3}^\dagger(t_0)\vert \Omega\rangle,     
\eeqa
where ${\bf R} :=({\bf x}_1 + {\bf x}_2 + {\bf x}_3)/3$, ${\bf r}:={\bf x}_1-{\bf x_2}$, ${\bf \rho}:={\bf x}_3 -({\bf x}_1 + {\bf x}_2 )/2$ are Jacobi coordinates,
and $J_3^\dagger(t_0)$ is a creation operator for three nucleons at time $t_0$.
The corresponding $R$-correlator is given by
\beqa
R_{J_{3}}^{3N}({\bf r}, {\bf \rho}, t)={F_{J_{3}}^{3N}({\bf r}, {\bf \rho}, t) \over (G_N(t))^3}.
\eeqa   
A genuine three-nucleon potential $V_{3NF} ({\bf r}, {\bf \rho})$ at the LO analysis is extracted from the time-dependent Schr\"odinger equation as
\beqa
\left[ -{\nabla_r^2\over 2\mu_r} - {\nabla_\rho^2\over 2\mu_\rho} +\sum_{i<j} V_{2N}({\bf r}_{ij}) +  V_{3NF} ({\bf r}, {\bf \rho}) \right]
R_{J_{3}}^{3N}({\bf r}, {\bf \rho}, t) = -{\partial \over \partial t} R_{J_{3}}^{3N}({\bf r}, {\bf \rho}, t),~~~~
\eeqa
where $V_{2N}({\bf r}_{ij})$ with ${\bf r}_{ij} := {\bf x}_i -{\bf x}_j$ denotes two-nucleon potential between $i$-th and $j$-th nucleons,
$\mu_r= m_N/2$, $\mu_\rho =2 m_N/3$ are reduced masses for corresponding Jacobi coordinates.
Here non-relativistic approximations are employed.

In lattice QCD calculations of three-nucleon potentials, there have been two big numerical obstructions.
One was a problem of enormous computational costs for  three-nucleon potentials.
This problem has been overcome by a novel computational algorithm, called a unified contraction algorithm (UCA)~\cite{Doi:2012xd},
which unifies two kinds of (Wick and color/spinor) contractions and then eliminates redundant calculations systematically. 
As a result, a factor of 192 speedup is achieved for calculations of three-nucleon potentials.
The other is  to handle huge degree of freedoms for  $V_{3NF} ({\bf r}, {\bf \rho})$, which is a factor of $L^3$ larger than $V_{2N}({\bf x}_{ij})$.
To avoid the factor of $L^3$, a special geometry of three nucleons has been employed in lattice QCD simulations  so far~\cite{Doi:2011gq,Aoki:2020bew}.
That is called a ``liner setup", where three nucleons are aligned linearly with equal spacing of $r_2:= \vert {\bf r}\vert/2$ and ${\bf \rho}=0$.
Furthermore,  quantum numbers of the system are restricted as $(I, J^P)=(1/2,1/2^+)$, the triton channel.
To reduce statistical fluctuations associated with subtractions of $V_{2N}$, $V_{3NF}$ has been calculated from a special channel,
given by ${1\over \sqrt{6}}[-p_\uparrow n_\uparrow n_\downarrow + p_\uparrow n_\downarrow n_\uparrow 
-  n_\uparrow n_\downarrow p_\uparrow + n_\downarrow n_\uparrow p_\uparrow + n_\uparrow p_\uparrow n_\downarrow - n_\downarrow p_\uparrow n_\uparrow]$.

\begin{figure}[h]
  \centering
    \includegraphics[width=0.49\textwidth]{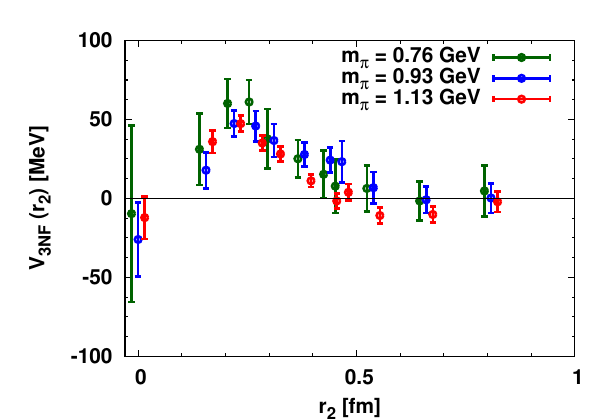}
  \includegraphics[width=0.49\textwidth]{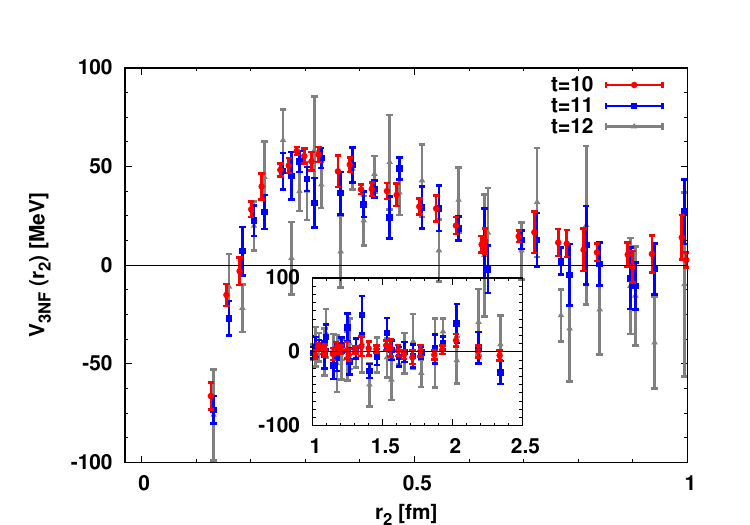}
  \caption{Three-nucleon potentials in the triton channel with the linear setup.
  (Left) Results from 2-flavor lattice QCD at $m_\pi=0.76, 0.93, 1.13$ GeV. 
    (Right) Results from (2+1) flavor QCD at $m_\pi=0.51$ GeV.
  Figures are taken from~\cite{Aoki:2020bew}.}
\label{fig:pot_3N} 
\end{figure}
The three-nucleon potential $V_{3NF}$ has been calculated in 2-flavor lattice QCD at $a\simeq 0.16$ fm on a $16^3\times 32$
with three different heavy pion masses, $(m_\pi, m_N) \simeq (0.76,1.61), (0.93,1.85)$, and $(1.13,2.15)$ GeV~\cite{Doi:2011gq,Aoki:2020bew}.
Fig.~\ref{fig:pot_3N} (Left) plots $V_{3NF}(r_2)$ at $m_\pi=0.76$ (green), 0.93 (blue), 1.13 (red) GeV.
While results at $r_2\le 0.2$ fm may suffer from lattice artifacts, a repulsive interaction at short distance, $r_2\simeq 0.2 - 0.4$ fm, is observed.
Repulsive short-ranged three-nucleon interactions are phenomenologically required to explain properties of high density nuclear matter.
At long distances, on the other hand, $V_{3NF}$ is negligibly small, probably due to heavier pion masses, which suppress
two-pion-exchanges responsible for three-nucleon interactions at long distances. 

Fig.~\ref{fig:pot_3N} (Right) shows $V_{3NF}(r_2)$ in $(2+1)$ flavor lattice QCD at $a\simeq 0.09$ fm on a $64^3\times 64$ lattice 
with a little lighter pion mass, $m_\pi\simeq 510$ MeV.
Extractions are made at three time separation, $t=10$ (red), 11 (blue) and 12 (black).
Except a very short distance region where lattice discretization errors could be substantial, the short-range repulsion is again observed 
in $V_{3NF}(r_2)$  at $r_2\simeq 0.2 - 0.7$ fm, which is wider than ones  in Fig.~\ref{fig:pot_3N} (Left) at heavier pion masses.
It is thus important to perform studies of three-nucleon potentials at lighter pion masses toward the physical pion mass. 
 
\section{\textit{4. Hyperon interactions}}
Potentials between hyperon-nucleon ($YN$) and hyperon-hyperon ($YY$) give a key to
understand nuclear many-body systems with strange quarks as impurity.
They are also essentially important to explore structures of a neutron star core, where a strangeness degree of freedom is expected to appear due to high density at the core.
Experimental data on $YN$ and $YY$ scatterings so far, however, are not sufficient to determine hyperon potentials, 
while spectroscopic studies of $\Lambda$ and $\Lambda\Lambda$ hypernuclei give some information on $\Lambda N$ and $\Lambda\Lambda$ interactions.  Recently, nucleus-nucleus collision experiments at RHIC and LHC also provide information on hyperon interactions.
In these circumstances, it is necessary to perform lattice QCD studies  in the HAL QCD method for first principle determination of hyperon interactions.

\subsection{4-1. Baryon interactions in the flavor SU(3) limit}
In lattice QCD simulations,   parameters in QCD such as quark masses can be varied from values in Nature.
To grasp  rough but global structures of hyperon interactions, it is more suitable to consider an idealized flavor SU(3) symmetric world, where
$u,d$ and $s$ quarks have a common non-zero mass.
In this limit, essential features of hyperon interaction can be captured without
effects associated with
quark mass differences.

A single baryon ground state belongs to the flavor octet with spin 1/2  in the flavor SU(3) limit, and
a state of two octet baryons with a  given angular momentum is labeled by an irreducible representation of the flavor SU(3) as
\beqa
{\bf 8} \otimes {\bf 8} = \underbrace{{\bf 27}\oplus {\bf 8}_s \oplus {\bf 1}}_{\rm symmetric} \oplus 
 \underbrace{\overline{\bf 10}\oplus {\bf 10} \oplus {\bf 8}_a}_{\rm anti-symmetric}, 
 \label{eq:B8B8}
\eeqa 
where ``symmetric" and ``anti-symmetric" represent symmetric properties under an exchange of  flavors for two octet baryons.
For a system with an orbital S-wave, a totally antisymmetric nature of fermions (Pauli principle)
implies {\bf 27}, {\bf 8}$_s$ and {\bf 1} to be spin-single ($^1S_0$) but  $\overline{\bf 10}$, {\bf 10} and {\bf 8}$_a$ to be spin-triplet ($^3S_1 - ^3D_1$),
where a subscript of the {\bf 8} representation  stands for the exchange symmetry.

\begin{figure}[h]
  \centering
  \includegraphics[width=0.32\textwidth]{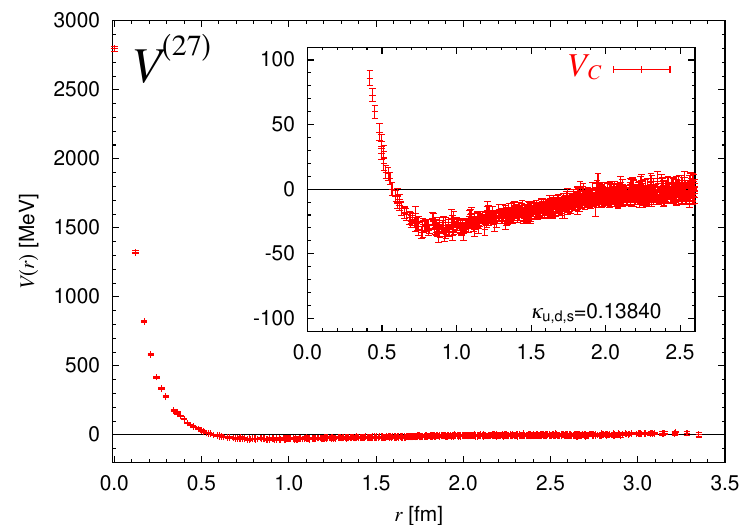}
  \includegraphics[width=0.32\textwidth]{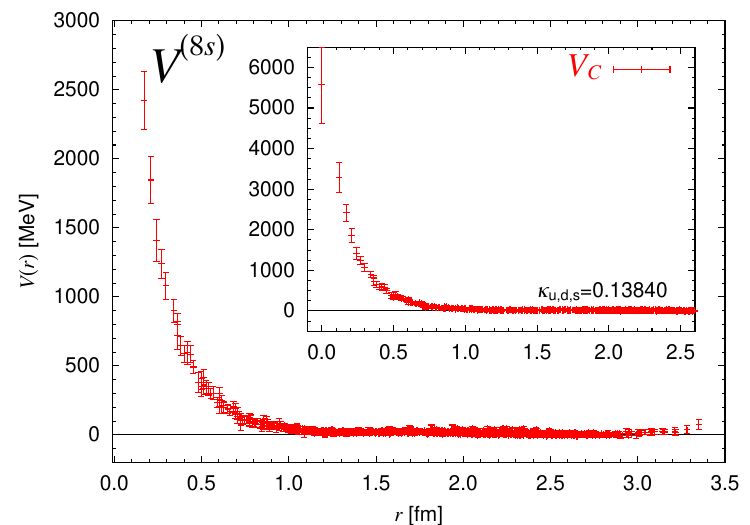}
  \includegraphics[width=0.32\textwidth]{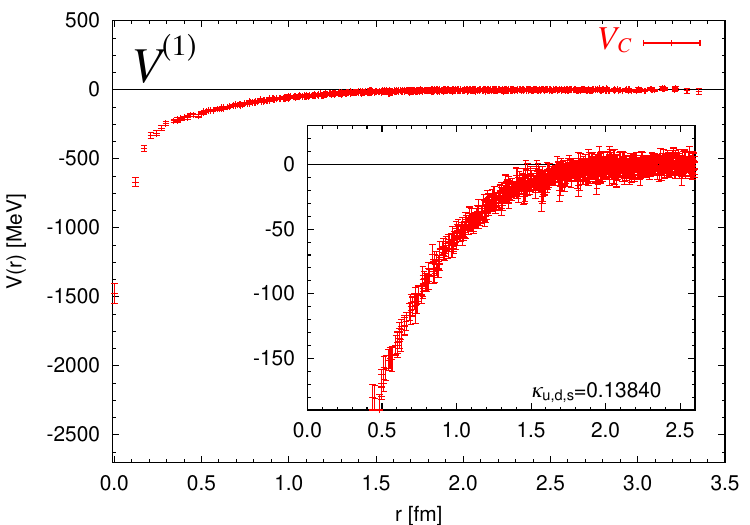}\\
\includegraphics[width=0.32\textwidth]{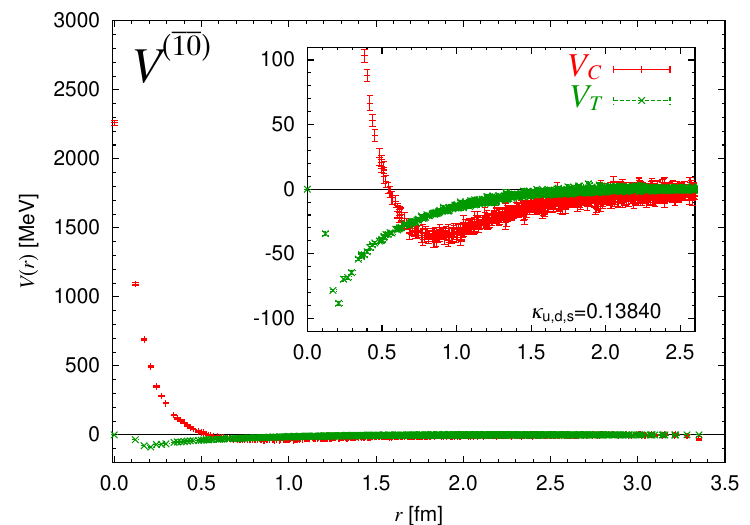}
  \includegraphics[width=0.32\textwidth]{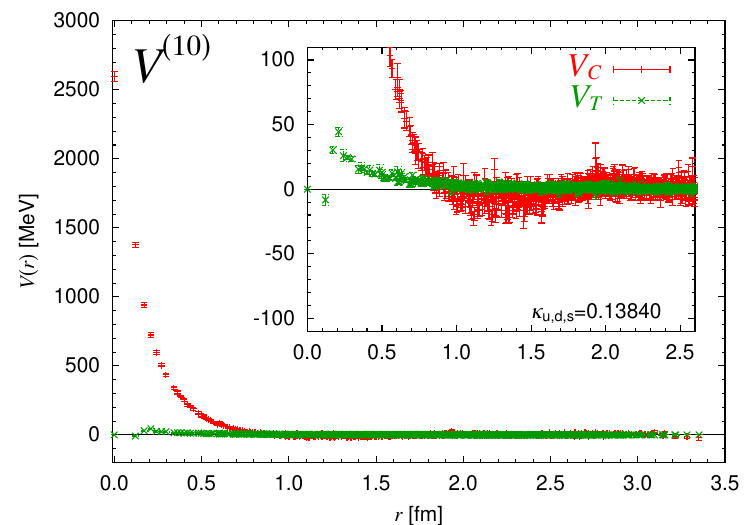}
  \includegraphics[width=0.32\textwidth]{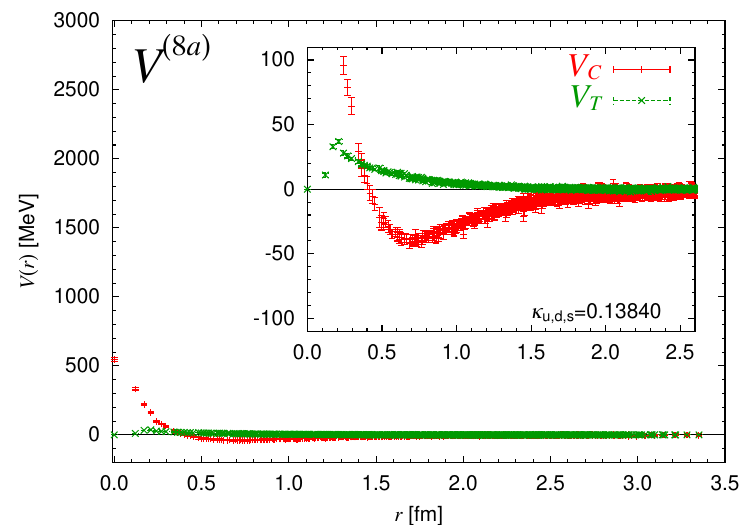}   
  \caption{(Upper)  $BB$ central potentials in {\bf 27} (Left), ${\bf 8}_s$ (Center) and {\bf 1} (Right) representations.
    (Lower) $BB$ central potentials (red) and tensor potentials (green) in  $\overline{\bf 10}$ (Left), {\bf 10} (Center) and ${\bf 8}_a$ (Right) representations. All potentials are extracted from lattice QCD simulations at $M_{\rm ps}\simeq 470$ MeV.
     Figures are  taken from~\cite{Inoue:2011ai}.
  }
\label{fig:pot_SU3} 
\end{figure}
Fig.~\ref{fig:pot_SU3} (Upper) shows central potentials in  {\bf 27} (Left), ${\bf 8}_s$ (Center) and {\bf 1} (Right) representations for the $^1S_0$ channel, while Fig.~\ref{fig:pot_SU3} (Lower) gives  central potentials (red) and tensor potentials (green) in $\overline{\bf 10}$ (Left), {\bf 10} (Center) and ${\bf 8}_a$ (Right) representations for the $^3S_1 - ^3D_1$ channel~\cite{Inoue:2011ai}.
These results are obtained using 3-flavor full QCD gauge ensembles generated on a $32^3\times 32$ lattice at $a\simeq 0.12$ fm 
with $(M_{\rm ps}, M_B) \simeq (469, 1161)$ MeV, where $M_{\rm ps}$ and $M_B$ are masses of a pseudo-scalar meson and an octet baryon, respectively. 
The central potential in the {\bf 27} representation (upper-left), $V_{\rm C}^{\bf (27)}(r)$, corresponds to the $NN$ potential in the $^1S_0$ channel, while $V_{\rm C}^{(\overline{\bf 10})}(r)$ (red) and $V_{\rm T}^{(\overline{\bf 10})}(r)$ (green) , the central and tensor potentials in  the $\overline{\bf 10}$ representation (lower-left), correspond to the $NN$ potentials in the $^3S_1-^3D_1$ channel. 
Both central potentials  have a repulsive core at short distance with an attractive pocket around 0.8 fm.
These features have already been observed in $NN$ potentials in previous section.
The central potential in the {\bf 10} representation (lower-center), $V_{\rm C}^{\bf (10)}(r)$ (red), has a stronger repulsive core and a weaker attractive pocket than   $V_{\rm C}^{\bf (27)}(r)$ and $V_{\rm C}^{(\overline{\bf 10})}(r)$.
The central potential in the ${\bf 8}_s$ representation (upper-center), $V_{\rm C}^{({\bf 8}_s)}(r)$, has the strongest repulsive core among all 6 representation without attraction, while the central potential in the ${\bf 8}_a$ representation  (lower-right),  $V_{\rm C}^{({\bf 8}_a)}(r)$ (red), has a weaker repulsive core and a deep attractive pocket. In contrast to all other representations, the central potential in the {\bf 1} representation (upper-right), $V_{\rm C}^{\bf (1)}(r)$, is attractive at all distances and has an attractive core instead of a repulsive core at short distance. 

Above features of potentials at short distances are consistent with predictions by the SU(6) quark model. In particular, the central potential in the ${\bf 8}_s$ representation is predicted to be strongly repulsive at short distance in the SU(6)  quark model~\cite{Oka:2000wj,Fujiwara:2006yh}, 
since six quarks cannot occupy the same orbital state due to the Pauli exclusion for quarks. 
The potential in the {\bf 1} representation, on the other hand, does not suffer from the  quark Pauli exclusion at all, so that
it can become attractive due to a short-range one gluon exchange. 
These agreements between the lattice result and the quark model on short distance behaviors suggest that the quark Pauli exclusion as well as the one-gluon-exchange play an essential role for repulsive cores in $BB$ interactions~\cite{Inoue:2010hs,Inoue:2011ai}.

\subsection{4-2. $H$ dibaryon in the flavor SU(3) limit}
\begin{figure}[h]
  \centering
  \includegraphics[width=0.49\textwidth]{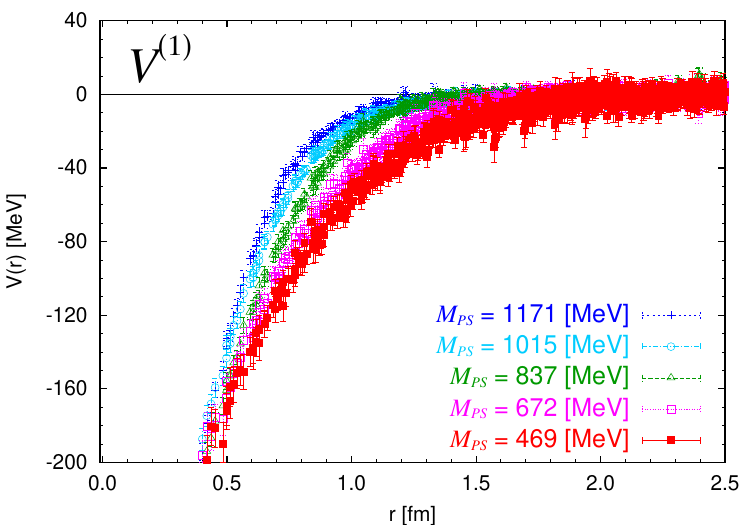}
  \includegraphics[width=0.49\textwidth]{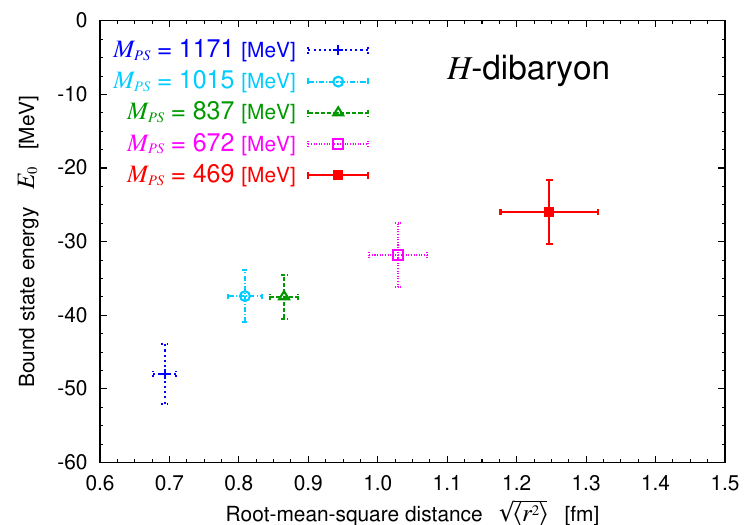}
  \caption{(Left)  The flavor singlet potential $V_{\rm C}^{({\bf 1})}(r)$ at  $M_{\rm ps}\simeq$ 1171 (blue), 1015 (cyan), 837 (green), 672 (magenta), 469 (red) MeV on a $32^3\times 32$ lattice in 3-flavor full QCD.
    (Right) The binding energy and the root-mean-square distance $\sqrt{ \langle r^2\rangle}$ of the bound $H$ dibaryon state 
    in the flavor singlet representation, where errors represent only statistical uncertainties. 
     Figures are  taken from~\cite{Inoue:2011ai}.
  }
\label{fig:pot_H} 
\end{figure}
The central potential in the the flavor singlet representation, $V_{\rm C}^{({\bf 1})}(r)$, shows an attractive behavior at all distances including very short distance, which may produce a bound state in this representation, called the $H$ dibaryon.
The attraction of $V_{\rm C}^{({\bf 1})}(r)$ becomes stronger toward the lighter $M_{\rm ps}$, as seen in Fig.~\ref{fig:pot_H} (Left).
These potentials are fitted by an attractive Gaussian core plus a long range (Yukawa)$^2$ with 5 parameters as
\beqa
V_{\rm C}^{({\bf 1})}(r) = b_1 e^{-b_2 r^2} + b_3 (1-e^{-b_4 r^2}) \left( {e^{-b_5 r}\over r}\right)^2.
\eeqa
Solving the Schr\"odinger equation with the fitted potential in an infinite volume, 
it is found that one bound state, the $H$ dibaryon, with binding energy of 20-50 MeV exists at each value of $M_{\rm ps}$~\cite{Inoue:2011ai,Inoue:2010es}. 
Fig.~\ref{fig:pot_H} (Right) shows the binding energy and the root-mean-square (rms) distance of the bound state at each $M_{\rm ps}$.
Despite a fact that the potential is more attractive as $M_{\rm ps}$ decreases, the binding energy of the $H$ dibaryon decreases in the current range of $M_{\rm ps}$, because the attraction is compensated by an increase of a kinetic energy for the lighter baryon mass.

Since the binding energy of the $H$ dibaryon is not so sensitive to the current range of $M_{\rm ps}$, and
is comparable to a splitting between hyperon masses in Nature, there may be a possibility of weakly bound or resonant $H$ dibaryon even in the real world with physical pion and kaon masses in the presence of the flavor SU(3) breaking.
To make a definite conclusion on a fate of $H$ dibaryon in Nature, however, $\Lambda\Lambda - N\Xi$ ( plus $\Sigma\Sigma$ if necessary) coupled channel analysis is required in the (2+1) flavor lattice QCD simulations, as will be discussed in the next subsection.

\subsection{4-3.  $\Lambda\Lambda - N\Xi$ interactions at  the almost physical point and the fate of the $H$-dibaryon}
With the flavor SU(3) breaking, the flavor singlet representation for two octet baryons is a coupled channel among  $\Lambda\Lambda$, $N\Xi$ and $\Sigma\Sigma$ states
in the S-wave with $I=0$ (isospin singlet), $S=0$ (spin singlet) and $s=-2$ (strangeness -2).

Such coupled channel analyses have been performed in lattice QCD simulations at heavy pion masses~\cite{Sasaki:2015ifa}, and 
the latest calculation~\cite{HALQCD:2019wsz} has been carried out
with an almost physical pion and kaon masses, $m_\pi\simeq 146$ MeV and $m_K\simeq 525$ MeV,
on a gauge ensemble 
in the (2+1)-flavor lattice QCD at $a=0.0846(7)$ fm on a $96^4$ lattice, corresponding to a spatial extension $L\simeq 8.1$ fm in the physical unit~\cite{Ishikawa:2015rho}.
Masses of octet baryons are $m_N\simeq  955$ MeV,  $m_\Lambda \simeq  1140$ MeV, $m_\Sigma \simeq  1222$ MeV and $m_\Xi \simeq  1355$ MeV. It turns out that the $\Sigma\Sigma$ state does not affect interactions in this channel at low energies, since
its threshold $m_\Sigma + m_\Sigma \simeq 2444$ MeV is much higher than the other two, $m_\Lambda+ m_\Lambda\simeq 2280$ MeV and $m_N + m_\Xi\simeq$ 2310 MeV. 

The leading order local $2\times 2$ coupled channel potentials between $\Lambda\Lambda$ and $N\Xi$ have been extracted 
from two octet baryon states
in the $I=0$, $^1S_0$ channel.
The $\Lambda\Lambda$ state in $^1S_0$ automatically has $I=0$, while the $N\Xi$ state in $^1S_0$ is projected to $I=0$.
The $\Lambda\Lambda$ and $N\Xi$ diagonal potentials  are well fitted by
\beqa
V^{\Lambda\Lambda}(r) &=&\sum_{i=1}^2 \alpha_i^{\Lambda\Lambda} e^{-{r^2\over (\beta_i^{\Lambda\Lambda})^2}} +\lambda_2^{\Lambda\Lambda}
\left[Y(\rho_2^{\Lambda\Lambda}, m_\pi, r)\right]^2, 
\label{eq:fit_LL}
\\
V^{N\Xi}(r) &=& \sum_{i=1}^3 \alpha_i^{N\Xi} e^{-{r^2\over (\beta_i^{N\Xi})^2}} +\lambda_2^{N\Xi}
\left[Y(\rho_2^{N\Xi}, m_\pi, r)\right]^2 +\lambda_1^{N\Xi}Y(\rho_1^{N\Xi}, m_\pi, r), ~~~~
\label{eq:fit_NX}
\eeqa 
while off-diagonal ones are found to be $V^{\Lambda\Lambda}_{N\Xi}(r) \simeq  V^{N\Xi}_{\Lambda\Lambda}(r)$ except at short distance, and thus its average,
$\bar V^{\Lambda\Lambda}_{N\Xi}(r) :=[V^{\Lambda\Lambda}_{N\Xi}(r) + V^{N\Xi}_{\Lambda\Lambda}(r) ]/2$,  is fitted for analysis as
\beqa
\bar V^{\Lambda\Lambda}_{N\Xi}(r) &=& \sum_{i=1}^2 \alpha_i e^{-{r^2\over \beta_i^2}} +\lambda_1
Y(\rho_1, m_K, r),
\label{eq:fit_nonD}
\eeqa 
where $Y$ is a Yukawa function with a from factor, given by
\beqa
Y(\rho, m,r) := \left( 1- e^{-{r^2\over \rho^2}}\right) {e^{- m r}\over r}, 
\eeqa
and $m_\pi = 146$ MeV and $m_K=525$ MeV are fixed for the above fit.
In eqs.~(\ref{eq:fit_LL}), (\ref{eq:fit_NX}) or (\ref{eq:fit_nonD}), Gaussian functions describe a short range part of the potential,  while
the Yukawa functions are motivated by a meson exchange contribution at medium and long distances:
the  Yukawa potential in eq.~(\ref{eq:fit_NX}) or eq.~(\ref{eq:fit_nonD}) represents the longest range single-pion or single-kaon exchange process
in the $N\Xi - N\Xi$ diagonal interaction or in the $\Lambda \Lambda - N\Xi$ transition, respectively, while
the Yukawa potential is absent in the $\Lambda\Lambda - \Lambda\Lambda$ interaction, whose long-range part does not exchange isospin and strangeness.
Instead, the squared Yukawa potential in eq.~(\ref{eq:fit_LL}) represents the two-pion exchange process, which
dominates at long distance in the $\Lambda\Lambda - \Lambda\Lambda$ interaction.

\begin{figure}[h]
  \centering
  \includegraphics[width=0.49\textwidth]{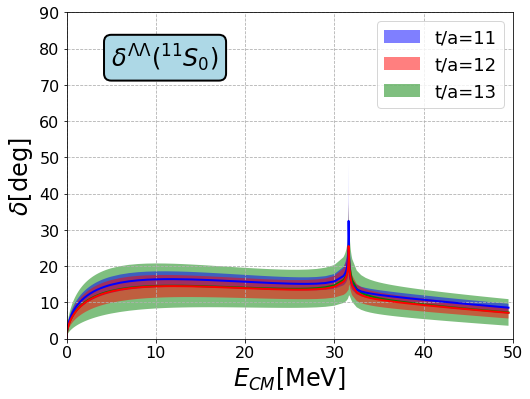}
  \includegraphics[width=0.49\textwidth]{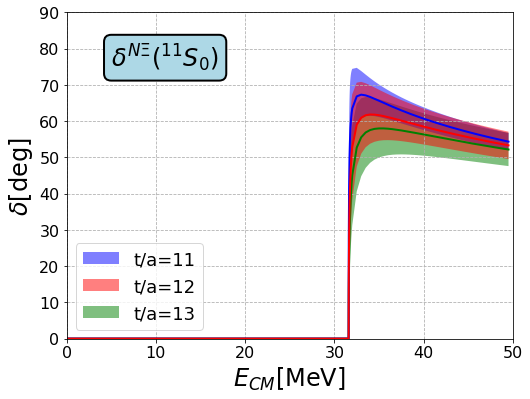}
  \caption{(Left)  The $\Lambda\Lambda$ scattering phase shift in the $^1S_0$ and $I=0$ channel as a function of the center-of-mass energy $E_{\rm CM} =k^2/m_\Lambda$ with $k$ being the relative momentum between $\Lambda\Lambda$.
    (Right) The $N\Xi$ scattering phase shift in the $^1S_0$ and $I=0$ channel as a function of $E_{\rm CM}$.
     Figures are  taken from~\cite{HALQCD:2019wsz}.
  }
\label{fig:phase_LL_NX} 
\end{figure}
Fig.~\ref{fig:phase_LL_NX} shows  $\Lambda\Lambda$ and $N\Xi$ scattering phase shifts in the $^1S_0$ and $I=0$ channel as a function of the center-of-mass energy $E_{\rm CM} =k^2/m_\Lambda$ with $k$ being the relative momentum between $\Lambda\Lambda$
for $t/a=11,12,13$ as time separations between source and sink operators. 
The $t$-dependence is minor within statistical errors, showing that the leading order local potential approximation is reasonably good.
 An attraction between $\Lambda\Lambda$ is rather weak, as seen in Fig.~\ref{fig:phase_LL_NX} (Left).
 Hence no bound or resonant dihyperon ($H$ dibaryon) exists around the $\Lambda\Lambda$ threshold in the (2+1) flavor QCD at the almost physical
 point. 
 A low energy part of the $\Lambda\Lambda$ phase shift in  Fig.~\ref{fig:phase_LL_NX} (Left) 
 is fitted by the S-wave effective range expansion (ERE) formula as 
 \beqa
 k \cot \delta_{\Lambda\Lambda}(k) = -{1\over a_0^{\Lambda\Lambda}} +{1\over 2} r_{\rm eff}^{\Lambda\Lambda} k^2 + O(k^4), 
 \eeqa
 where the sign convention of $a_0$ in nuclear and atomic physics is employed.
 The fit provides  scattering phase shift and effective range, $a_0^{\Lambda\Lambda} =-0.8(3)$ fm and
 $ r_{\rm eff} ^{\Lambda\Lambda} = 5.5(1.0)$ fm.
 which are confirmed to be consistent with a constraint from the $\Lambda\Lambda$ momentum correlation of p-p and p-Pb collisions~\cite{ALICE:2018ysd,ALICE:2019eol}, whose sign convention of $a_0$, however, is opposite to the above.
 
The $\Lambda\Lambda$ phase shift in Fig.~\ref{fig:phase_LL_NX} (Left) shows a sharp enhancement followed by a rapid drop near the $N\Xi$ threshold, due to the off-diagonal coupling.  Correspondingly, Fig.~\ref{fig:phase_LL_NX} (Right) shows a sharp increase of the $N\Xi$ phase shift $\delta^{N\Xi}(k)$ up to about $60^\circ$ just above the $N\Xi$ threshold, which is a consequence of the significant $N\Xi$ attraction in the $^1S_0$ with $I=0$.
The $N\Xi$ system in this channel is in the unitary region and a virtual pole is created, as in the case of a virtual pole of the $NN$ system in $^1S_0$ channel. Thus, the analysis by the coupled channel HAL QCD potentials indicates that the $H$ dibaryon appears as a virtual state of the $N\Xi$ in the  $^1S_0$ ($I=0$) channel at the almost physical point, $m_\pi\simeq 146$ MeV and $m_K\simeq 525$ MeV.
  For the detailed analysis of the virtual pole of $H$-dibaryon
  in the Riemann sheets of the coupled channel system,
  see Ref.~\cite{Kamiya:2021hdb}.

\subsection{4-4. $N\Xi$ interactions at the almost physical point}
The $N\Xi$ interactions in the S-wave can be classified as
$^{2I+1, 2s+1}S_J$ where $I=0,1$ denotes total isospin, $s=0,1$ total spin,
and $J$ total angular momentum,
and all four channels have been calculated
at the almost physical point~\cite{HALQCD:2019wsz}.
The coupled channel analysis  in the $^{11}S_0$ channel are already given in the previous subsection,
and the results in other three channels
are presented here.

The leading order central potentials for $N\Xi$ are extracted
from a single channel analysis in each $^{31}S_0, ^{13}S_1, ^{33}S_1$ channel.
Note that there exist channel couplings only from higher channels
as $\Lambda\Sigma, \Sigma\Sigma$, and the coupled channel
effect is effectively included in the obtained $N\Xi$ potentials.

The potentials are fitted using the functional form given in eq.~(\ref{eq:fit_NX}),
where the conditions
$
\lambda_1^{N\Xi}(^{11}S_0)
= -3 \lambda_1^{N\Xi}(^{31}S_0)
= -3 \lambda_1^{N\Xi}(^{13}S_1)
= 9  \lambda_1^{N\Xi}(^{33}S_1)
$
and
$
\lambda_2^{N\Xi}(^{11}S_0)
= \lambda_2^{N\Xi}(^{31}S_0)
= \lambda_2^{N\Xi}(^{13}S_1)
= \lambda_2^{N\Xi}(^{33}S_1)
$
are imposed based on the one-pion and two-pion exchange picture.
The range parameters $\beta_{1,2,3}^{N\Xi}$ and $\rho_{1,2}^{N\Xi}$
are also assumed to the same among different channels.
Scattering phase shifts are then obtained
by solving the Schr\"odinger equation with the fitted potential.

\begin{figure}[h]
  \centering
  \includegraphics[width=0.32\textwidth]{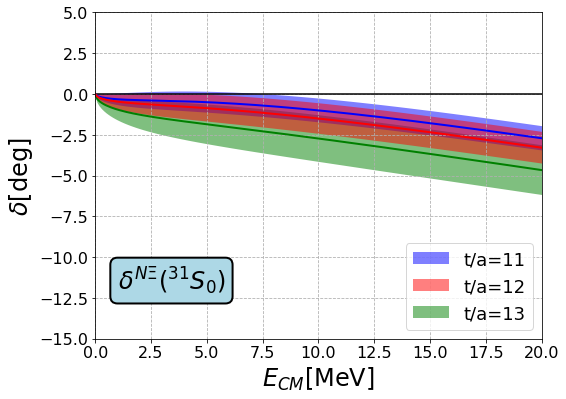}
  \includegraphics[width=0.32\textwidth]{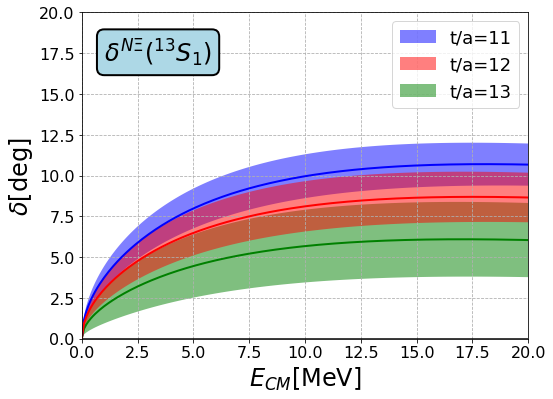}
  \includegraphics[width=0.32\textwidth]{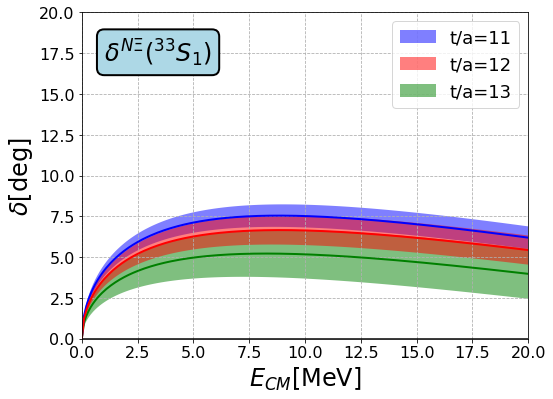}
  \caption{
    $N \Xi$ scattering phase shifts in ${^{31}S_0}$ (Left), ${^{13}S_1}$ (Center)
    and ${^{33}S_1}$ (Right) channels.
    Figures are  taken from~\cite{HALQCD:2019wsz}.
  }
 \label{fig:phase_NX}
\end{figure}

Shown in Fig.~\ref{fig:phase_NX} are scattering phase shifts
in ${^{31}S_0}$ (Left), ${^{13}S_1}$ (Center) and ${^{33}S_1}$ (Right) channels
at $t/a = 11, 12, 13$
as a function of the center-of-mass energy, $E_{\rm CM} = k^2 \times (1/(2m_N) + 1/(2m_\Xi))$.
The $t$-dependence is minor within statistical errors,
indicating that systematic errors associated with the derivative expansion
and inelastic state contaminations are small.
The obtained phase shifts in Fig.~\ref{fig:phase_NX}
as well as Fig.~\ref{fig:phase_LL_NX} (Right) 
show that
$N\Xi$ interactions in 
${^{31}S_0}$,
${^{13}S_1}$,
${^{33}S_1}$ and
${^{11}S_0}$
channels are
weakly repulsive,
weakly attractive,
weakly attractive
and
moderately attractive,
respectively.

These results can be used to predict two-baryon correlations
in p-p and p-Pb collisions.
The femtoscopic analysis for (spin-isospin averaged) $p$-$\Xi^-$ correlations
by ALICE Collaboration at LHC
show good agreement between
observations and
predictions based on HAL QCD $N\Xi$ interactions
with the Coulomb attraction taken into account~\cite{ALICE:2019hdt,ALICE:2020mfd}.
It is also possible to predict what kind of $\Xi$-hypernuclei exist,
using lattice QCD $N\Xi$ potentials.
In Refs.~\cite{Hiyama:2019kpw,Hiyama:2022jqh},
such few-body calculations 
are performed 
by a high precision variational approach,
the Gaussian Expansion Method,
using HAL QCD $N\Xi$ interactions for which a small extrapolation
from the almost physical point ($(m_\pi, m_K) = (146, 525)$ MeV)
to the physical point ($(m_\pi, m_K) = (138, 496)$ MeV) are performed.
It is found that
the $\Xi NNN$ system in $I=0, J^P = 1^+$ channel is bound
as the lightest $\Xi$-hypernucleus.
In addition, $\Xi N \alpha\alpha$ systems are found to have
a pair of bound states having $J^P = 1^-$ and $2^-$, called a spin-doublet bound state, 
in both $I=0, 1$ channels.
Interestingly, while the $1^-$ state has a  larger binding energy than the $2^-$ state in the $I=0$ channel,
this ordering is reversed in the $I=1$ channel. 
The opposite ordering of  the $1^-$ - $2^-$ spin-doublet state spectra between $I=0$ and $1$ is found to be
strongly correlated with relative strengths of $N\Xi$ interactions
in $^{11}S_0$, $^{13}S_1$, $^{31}S_0$ and $^{33}S_1$ channels.
Therefore, experimental search of these doublets by, e.g.,
$(K^-, K^+)$ and $(K^-, K^0)$ reactions on the $^{10}$B target,
will be important to confirm the lattice QCD prediction on the spin-isospin dependence
of $N\Xi$ interactions.

\section{\textit{5. Dibaryons at the almost physical point}}
A dibaryon is a bound state or resonance having a baryon number $B=2$, such as a deuteron, made of a proton and a neutron, which is only a stable dibaryon observed in Nature so far. A dibaryon can be classified in the flavor SU(3) representation as in eq.~(\ref{eq:B8B8}) for octet-octet baryons, where
the deuteron belongs to the $\overline{\bf 10}$ representation while the $H$ dibaryon was predicted to appear in the {\bf 1} representation~\cite{Jaffe:1976yi} but is likely to be a virtual state, as discussed in the previous section.
Classifications including decuplet ({\bf 10}) baryons are
\beqa
{\bf 10}\otimes {\bf 8} = \bf 35 \oplus {\bf 8} \oplus {\bf 10} \oplus {\bf 27} 
\eeqa
where $N\Omega$ and $N\Delta$ dibaryons were predicted in {\bf 8} and {\bf 27} representations~\cite{Goldman:1987ma,Oka:1988yq,Dyson:1964xwa}, respectively, and 
\beqa
{\bf 10}\otimes {\bf 10} = {\bf 28} \oplus {\bf 27} \oplus {\bf 35} \oplus \overline{\bf 10}, 
\eeqa
where $\Omega\Omega$ was predicted in the {\bf 28} representation~\cite{Zhang:1997ny} while $\Delta\Delta$ was predicted in the $\overline{\bf 10}$ representation~\cite{Dyson:1964xwa,Kamae:1976at},
whose candidate $d^*(2380)$ has been observed~\cite{WASA-at-COSY:2011bjg}.
Note that all decuplet single baryons are unstable against strong decays except for $\Omega$. 

In this section, a summary is given for results on dibaryons including  a charmed dibaryon by the HAL QCD potential method 
using the (2+1) full QCD ensemble at the almost physical point, whose properties such as hadron masses were already explained in Sec. 4-3.

\subsection{5-1. The most strange dibaryon $\Omega_{sss}\Omega_{sss}$}
Interactions between $\Omega_{sss}^-$ $\Omega_{sss}^-$  in the $^1S_0$ channel,
which belongs to the {\bf 28} representation,
 have been investigated at a heavy pion mass~\cite{HALQCD:2015qmg}, and recently at the almost physical point~\cite{Gongyo:2017fjb}.
 Here superscript $-$ of $\Omega_{sss}^-$ indicates its charge.
 
 \begin{figure}[h]
  \centering
  \includegraphics[width=0.49\textwidth]{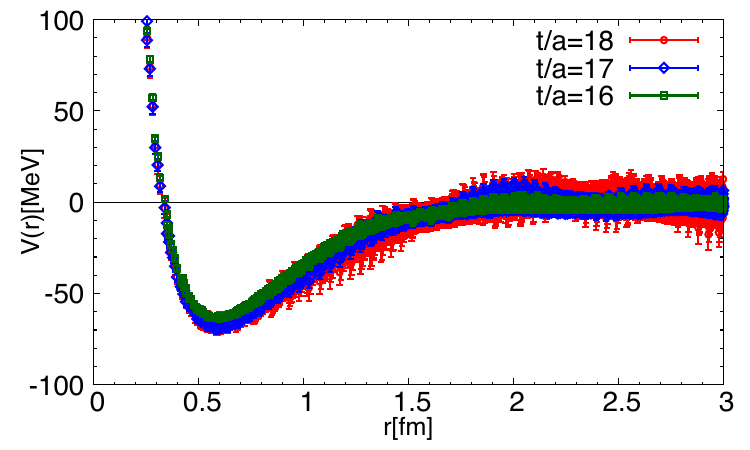}
  \includegraphics[width=0.49\textwidth]{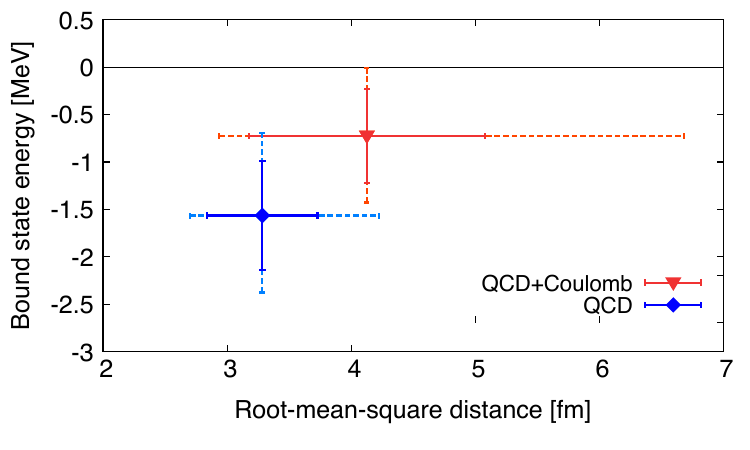}
  \caption{(Left)  The $\Omega^-_{sss}\Omega^-_{sss}$ potential in the $^1S_0$ channel at $t/a = 16$ (green), 17 (blue), 18 (red) as a function of $r$ [fm], obtained in (2+1) lattice QCD simulations at the almost physical point.
    (Right) The binding energy of the $\Omega^-_{sss}\Omega^-_{sss}$ dibaryon and its root-mean-square distance with and without
    the Coulomb repulsion (red solid triangle and blue solid diamond, respectively).  
     Figures are  taken from~\cite{Gongyo:2017fjb}.
  }
\label{fig:OmegaOmega} 
\end{figure}
 Fig.~\ref{fig:OmegaOmega} (Left) shows the $\Omega^-_{sss}\Omega^-_{sss}$ potential in the $^1S_0$ channel at $t/a=16,17,18$,
 obtained in (2+1) lattice QCD at  the almost physical point, 
 where $t$ is the time-separation between source and sink operators used to extract the potential.
While  the $\Omega^-_{sss}\Omega^-_{sss}$ potential has qualitative features similar to the central potentials for $NN$,
its repulsion is weaker and attraction is shorter ranged than $NN$ cases.
The  $\Omega^-_{sss}\Omega^-_{sss}$ potential allows one shallow bound state, whose binding energy is given by
$B_{\Omega^-_{sss}\Omega^-_{sss}} =$ 1.6(9) MeV in QCD without the Coulomb repulsion and 0.7(7) MeV with the Coulomb repulsion
between $\Omega^-_{sss}\Omega^-_{sss}$ as $\alpha_e/r$ with $\alpha_e :=e^2/(4\pi) = 1/137.036$.
This bound state may be called the most ``strange" dibaryon~\cite{Gongyo:2017fjb}, as its baryon number $B=2$ is carried by six strange quarks.
Fig.~\ref{fig:OmegaOmega} (Right) represents the binding energy and the root-mean-square distance in QCD + Coulomb (red solid triangle) and
in QCD (blue solid diamond).
A scattering length $a_0$ and an effective range $r_{\rm eff}$  in QCD are also extracted as
\beqa
a_0^{\Omega^-_{sss}\Omega^-_{sss}} = 4.6(1.3) {\rm fm}, \quad r_{\rm eff}^{\Omega^-_{sss}\Omega^-_{sss}} = 1.27(7) {\rm fm}, 
\eeqa 
 which indicates that the $\Omega^-_{sss}\Omega^-_{sss}$ in the $^1S_0$ channel is located near unitary region as
 $r_{\rm eff}/a_0 \simeq  0.28$,  similar order to $r_{\rm eff}/a_0 \simeq  0.32$ for $NN(^3S_1)$ (deuteron) and   $r_{\rm eff}/a_0 \simeq -0.12$ for
 $NN(^1S_0)$.
 
 The bound $\Omega^-_{sss}\Omega^-_{sss}$ can be best searched experimentally by two particle correlations in future relativistic nucleus-nucleus collisions~\cite{Morita:2019rph}.
 
\subsection{5-2. The most charming dibaryon $\Omega_{ccc}\Omega_{ccc}$}
Using the same gauge ensemble at the almost physical point, $\Omega_{ccc}^{++}\Omega_{ccc}^{++}$ interactions in $^1S_0$ channel 
have been investigated~\cite{Lyu:2021qsh,Lyu:2021cte}, where $\Omega_{ccc}^{++}$ is a decoupled baryon with a charge $++$ made of three charm quarks.
In this study, a charm quark has been treated as a valence quark without effects of dynamical charm quarks.
The charm quark mass has been interpolated to reproduce an experimental value of a spin averaged $1S$ charmonium mass,
$(m_{\eta_c} + 3 m_{J/\Psi} )/4 =$ 3068.5 MeV.
While no experimental value of $\Omega_{ccc}^{++} $ mass is available, 
$m_{\Omega_{ccc}^{++}} \simeq 4796(1)$ MeV at this charm quark mass is consistent with  $4789(22)$ MeV obtained in other lattice calculations~\cite{PACS-CS:2013vie}.

 \begin{figure}[h]
  \centering
  \includegraphics[width=0.49\textwidth]{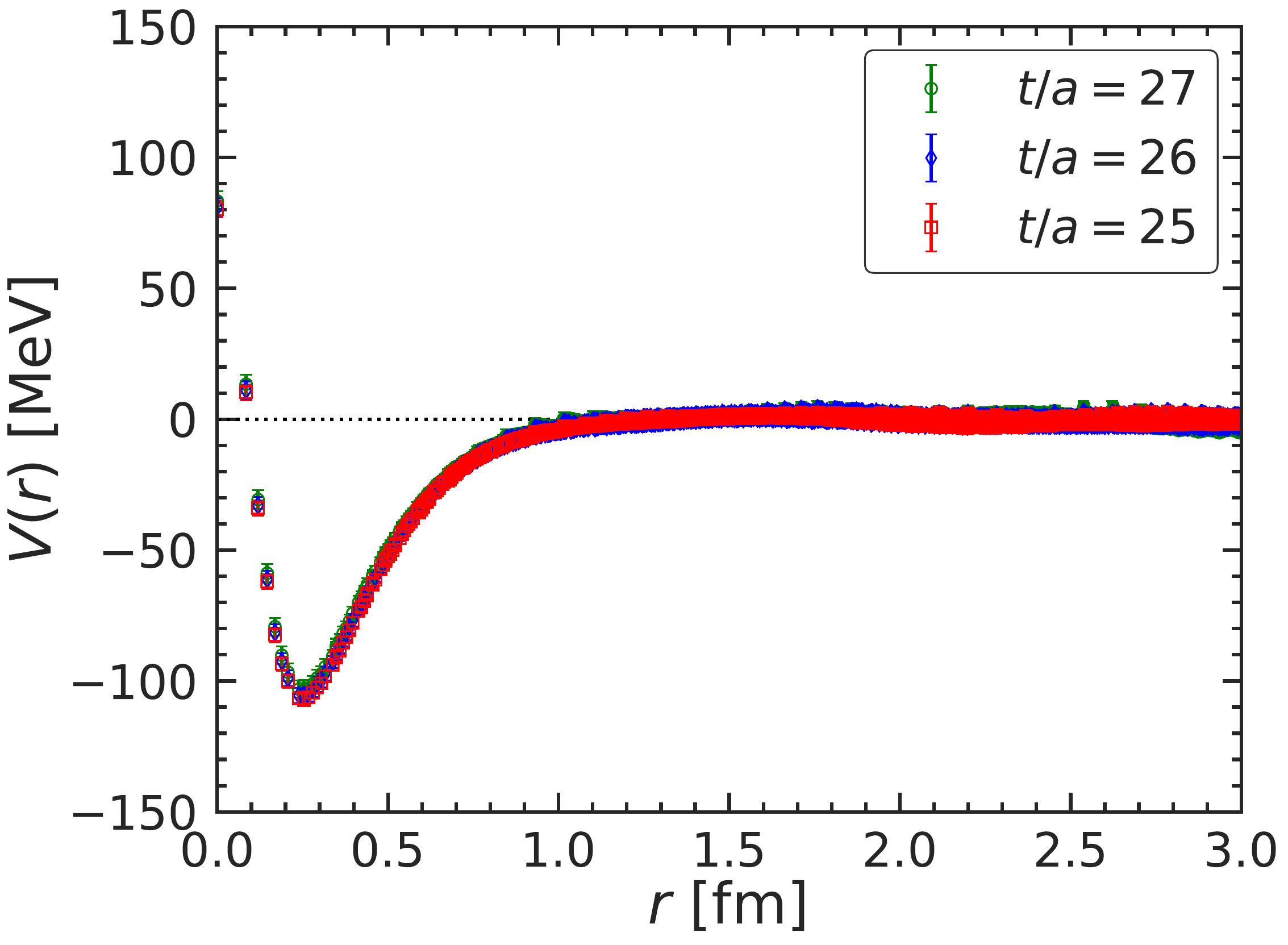}
  \includegraphics[width=0.49\textwidth]{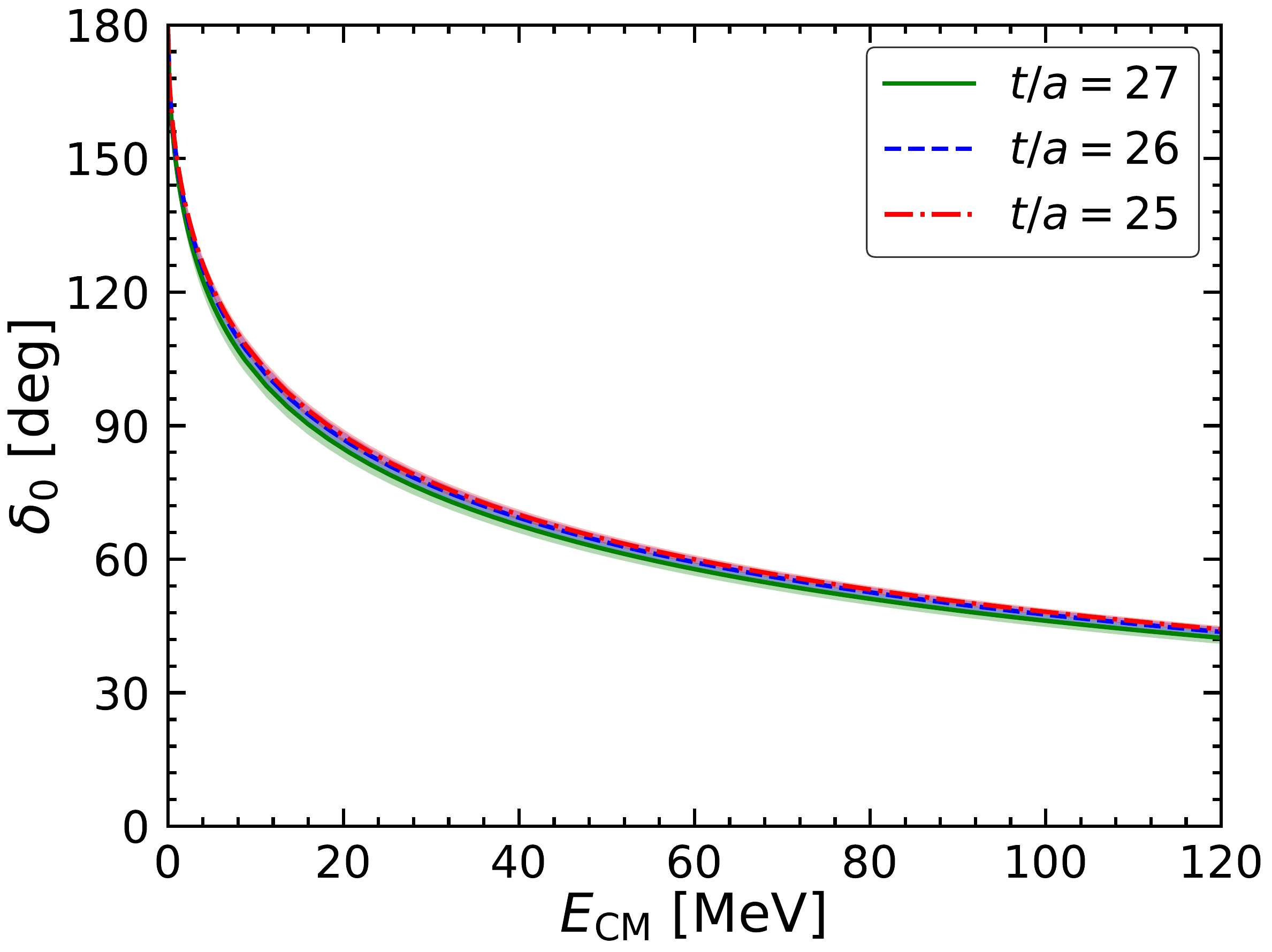}
  \caption{(Left)  The $\Omega^{++}_{ccc}\Omega^{++}_{ccc}$ potential in the $^1S_0$ channel at $t/a = 25$ (red squares), 26 (blue diamonds), 27 (green circles) as a function of separation $r$ [fm], obtained in (2+1) lattice QCD simulations at the almost physical point.  
    (Right)  The $\Omega^{++}_{ccc}\Omega^{++}_{ccc}$ scattering phase shift $\delta_0$ in the $^1S_0$ channel as a function of the center of mass kinetic energy $E_{\rm CM}$. 
     Figures are  taken from~\cite{Lyu:2021qsh}.
  }
\label{fig:Omega_ccc} 
\end{figure}
The $\Omega^{++}_{ccc}\Omega^{++}_{ccc}$ potential in the $^1S_0$ channel is plotted as a function of
separation $r$ between two baryons in Fig.~\ref{fig:Omega_ccc} (Left)
at $t/a= 25$ (red squares), 26 (blue diamonds), 27 (green circles).
As before, a small $t$ dependence indicates that systematic errors due to inelastic states and higher order terms in the derivative expansion
are comparable to or even smaller than statistical errors.

The $\Omega^{++}_{ccc}\Omega^{++}_{ccc}$ potential is repulsive at short distance but attractive at middle and long  distances,
and these features are qualitatively similar to $NN$ potentials and the  $\Omega^-_{sss}\Omega^-_{sss}$ potential, as already seen. 
Fig.~\ref{fig:Omega_ccc} (Right) shows the $\Omega^{++}_{ccc}\Omega^{++}_{ccc}$ scattering phase shift in the $^1S_0$ channel calculated by solving the Schr\"odinger equation with the potential in Fig.~\ref{fig:Omega_ccc} (Left),
as a function of the relativistic kinetic energy defined by $E_{\rm CM} = 2\sqrt{k^2 + m_{\Omega^{++}_{ccc}}^2}- 2 m_{\Omega^{++}_{ccc}}$
with a center of mass momentum $k$.
The phase shift starts from $180^\circ$,  indicating an existence of one bound state in the $\Omega_{ccc}^{++}\Omega_{ccc}^{++}$ system
(the most charming dibaryon~\cite{Lyu:2021cte}) in QCD without Coulomb repulsion,
whose binding energy is calculated as $B_{\Omega_{ccc}^{++}\Omega_{ccc}^{++}} = 5.7(1.3)$ MeV. 
The low-energy scattering parameters are extracted by the ERE as
\beqa
 a_0^{\Omega_{ccc}^{++}\Omega_{ccc}^{++}}= 1.6(1)~{\rm fm},\quad
  r_{\mathrm{eff}}^{\Omega_{ccc}^{++}\Omega_{ccc}^{++}}= 0.57(2)~{\rm fm}.
\eeqa

A Coulomb repulsion between two $\Omega_{ccc}^{++}$'s is estimated as
\beqa
V^{\rm Coulomb}(r) &=& \alpha_e \int\int d^3r_1d^3r_2 {\rho( \vert{\bf r}_1\vert) \rho (\vert{\bf r}_2 -{\bf r}\vert)\over \vert {\bf r}_1 -{\bf r}_2\vert}, 
\label{eq:Coulomb}
\eeqa
where a charge distribution inside the $\Omega_{ccc}^{++}$ is taken as
\beqa
\rho( r) = {12\sqrt{6}\over \pi r_d^3}e^{-2\sqrt{6} r/r_d}
\eeqa
 with a charge radius $r_d =0.410(6)$ fm~\cite{Can:2015exa}.
 The ERE with the Coulomb repulsion reads
 \beqa
 k \left[C_\eta^2 \cot \delta_0^{\rm C}(k) + 2\eta h(\eta) \right] = -{1\over a_0^{\rm C}} + {1\over 2} r_{\rm eff}^{\rm C} k^2 +O(k^4),
 \eeqa
where $\delta^{\rm C}_0(k)$ is a phase shift in the presence of Coulomb interactions, $C^2_\eta=\frac{2\pi\eta}{e^{2\pi\eta}-1}$, $\eta={2\alpha_e m_{\Omega_{ccc}^{++}}}/{k}$, $h(\eta)={\rm Re}[\Psi(i\eta)]-\ln(\eta)$, and $\Psi$ is the digamma function~\cite{Burke2011}.
Due to a large cancellation between the attraction in QCD and the Coulomb repulsion,
the bound state disappears and scattering parameters become
\beqa
a_0^{\rm C} = -19(10)~{\rm fm}, \quad r_{\rm eff}^{\rm C} =0.45(1)~{\rm fm}, 
\eeqa
whose ratio $r^{\rm C}_{\rm eff}/a^{\rm C}_0\simeq -0.024$ is considerably smaller in magnitude than $r_{\rm eff}/a_0\simeq -0.12$
for $NN (^1S_0)$.

\subsection{5-3. Comparisons of two systems}
In this subsections, some comparisons are made 
between $\Omega_{sss}^-$ $\Omega_{sss}^-$ and $\Omega^{++}_{ccc}\Omega^{++}_{ccc}$ systems.

 \begin{figure}[h]
  \centering
  \includegraphics[width=0.49\textwidth]{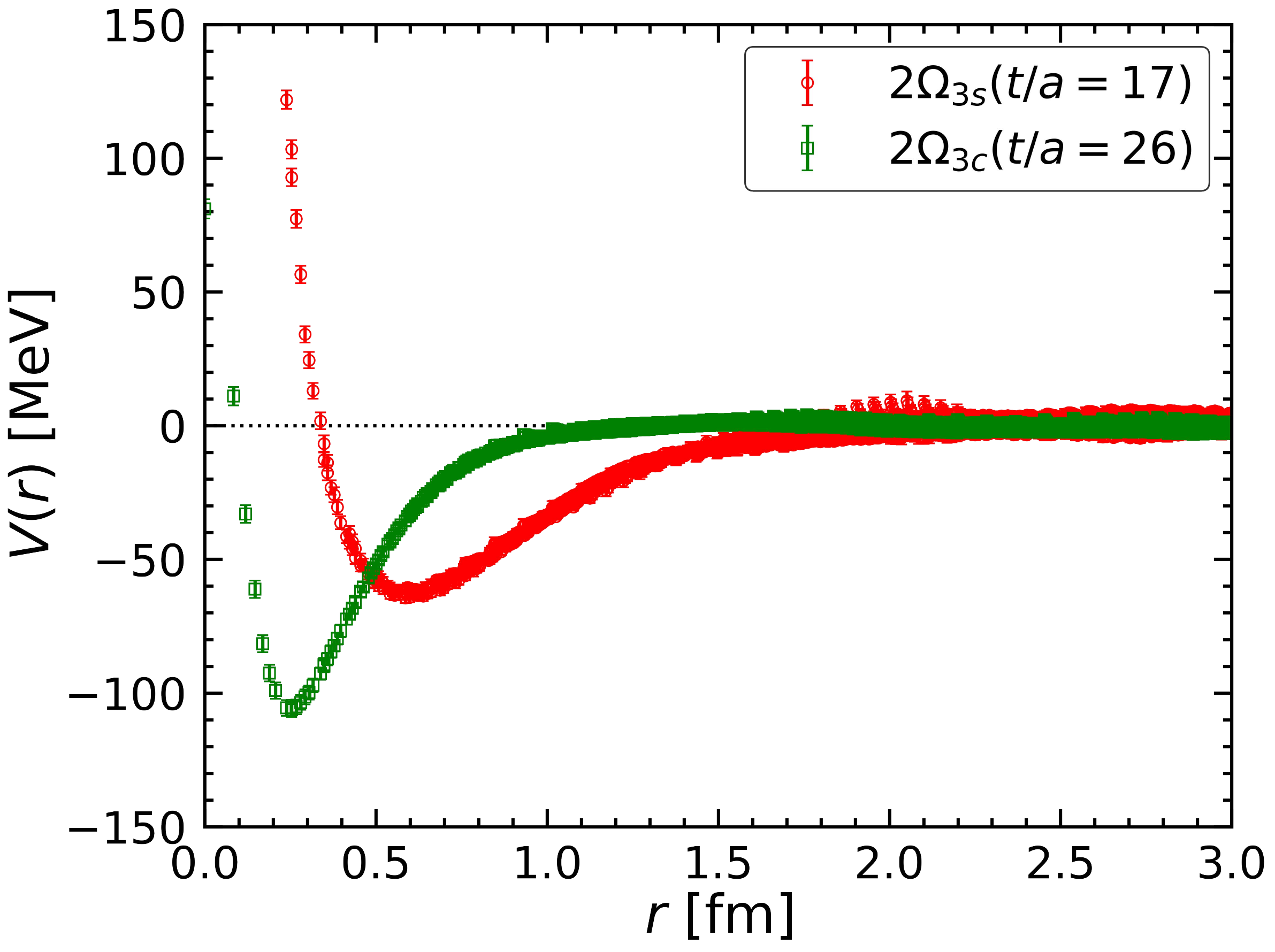}
  \includegraphics[width=0.49\textwidth]{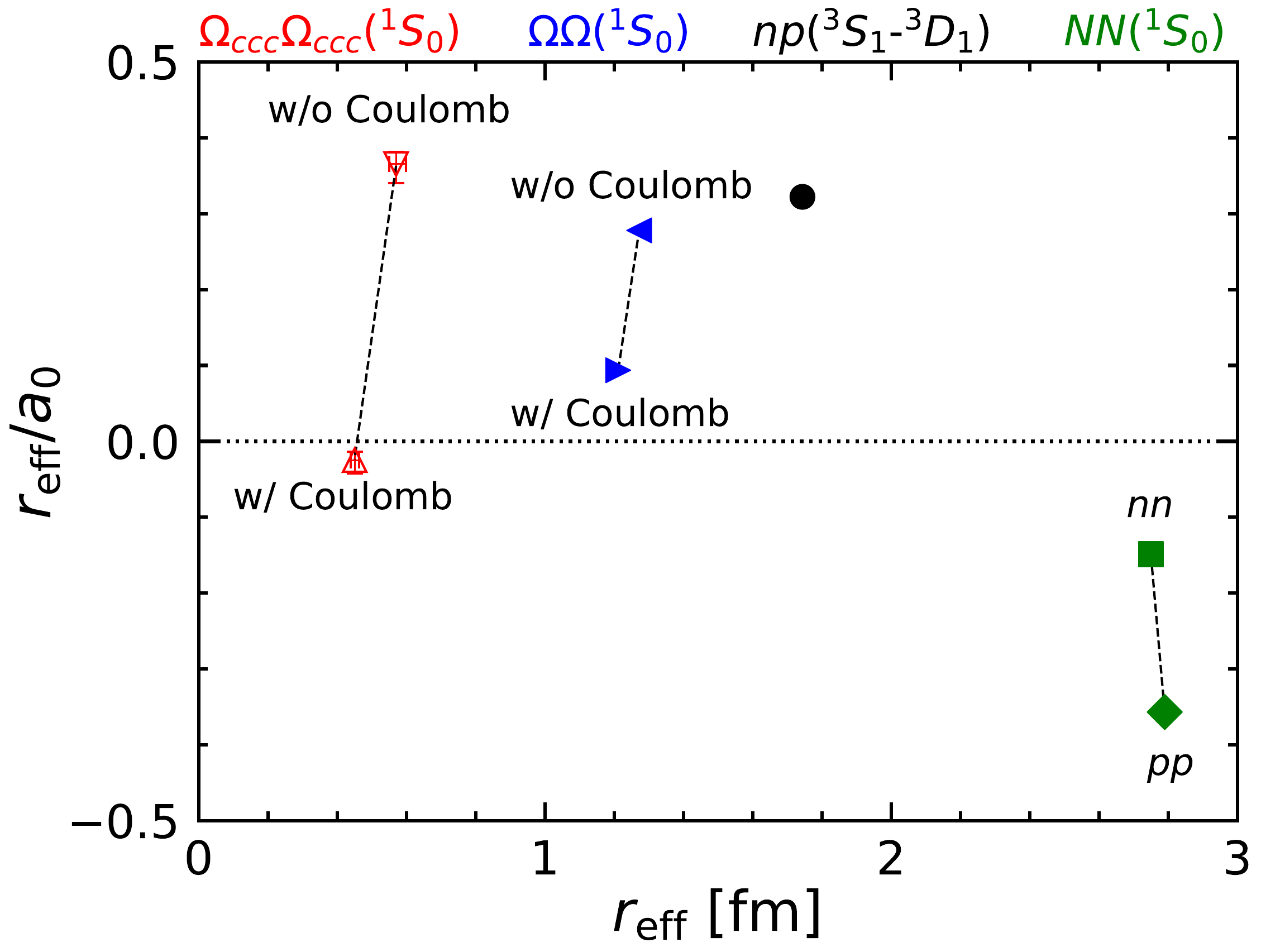}
  \caption{(Left)  The $\Omega_{sss}^-\Omega_{sss}^-$ potential at $t/a=17$ (red circles) and the 
  $\Omega^{++}_{ccc}\Omega^{++}_{ccc}$ potential  at $t/a = 26$ (green squares) as a function of separation $r$ [fm]. 
  Figure is  taken from~\cite{Lyu:2022tsd}.
    (Right)  A dimensionless ratio $r_{\rm eff}/a_0$ and $r_{\rm eff}$, where $a_0$ and $r_{\rm eff}$ are the scattering length and the effective range, respectively.
A red up(down) triangle and a blue right(left)  triangle correspond to  $\Omega^{++}_{ccc}\Omega^{++}_{ccc}$  and  $\Omega_{sss}^-\Omega_{sss}^-$   systems in the $^1S_0$ channel with (without) the Coulomb repulsion, respectively. 
A black circle represents the $NN$ system in the $^3S_1 - ^3D_1$ channel, while green square ($nn$) and diamond ($pp$) correspond to
the $NN$ system in the $^1S_0$ channel.  
     Figure is  taken from~\cite{Lyu:2021qsh}.
  }
\label{fig:Omega_comp} 
\end{figure}
Fig.~\ref{fig:Omega_comp} (Left) gives a comparison of potentials in the $^1S_0$ channel between   $\Omega_{sss}^- \Omega_{sss}^-$ (red) and  
$\Omega^{++}_{ccc}\Omega^{++}_{ccc}$ (green) systems.
First of all, a height of the $\Omega^{++}_{ccc}\Omega^{++}_{ccc}$ potential in the repulsive region $ r < 0.25$ fm 
is an order of magnitude smaller than that of the  $\Omega_{sss}^- \Omega_{sss}^-$ potential at its repulsive region $r< 0.5$ fm.
This difference may be qualitatively explained by the phenomenological quark model~\cite{Oka:2000wj}, as already seen at short distance parts of octet-octet potentials in the flavor SU(3) limit.
Since the color-magnetic (cm) interaction between  constituent quarks is proportional to a constituent quark mass inverse squared, 
it is qualitatively expected that $V_{\rm cm}^{cc}/V_{\rm cm}^{ss} \simeq ( m_s^*/m_c^*)^2 \sim (500/1500)^2 \sim 0.1$,
where $V_{\rm cm}^{ff^\prime}$ is a potential induced by the color-magnetic interaction between flavors $f$ and $f^\prime$ with the constituent quark mass  $m_f^*$.  
At medium and long distances,  on the other hand, exchanges of charmed mesons or strange mesons may dominate 
in  the $\Omega^{++}_{ccc}\Omega^{++}_{ccc}$ potential or  the $\Omega^{-}_{sss}\Omega^{-}_{sss}$ potential, respectively.
Thus mass differences between charmed mesons and strange mesons may explain 
a difference of a range in the attractive part of the potential,
$ 0.25\, {\rm fm} < r < 1.0\, {\rm fm}$  in the $\Omega^{++}_{ccc}\Omega^{++}_{ccc}$ potential and
$0.5\,   {\rm fm} < r < 2.0\, {\rm fm}$ in the $\Omega^{-}_{sss}\Omega^{-}_{sss}$ potential.
As already mentioned,  due to the cancellation between the attraction at medium-to-long distance and the repulsion at short distance,
only one loosely bound state appears in each of  $\Omega_{sss}^-\Omega_{sss}^-$ and $\Omega^{++}_{ccc}\Omega^{++}_{ccc}$
without the Coulomb repulsion.

In Fig.~\ref{fig:Omega_comp} (Right), a dimensionless ratio $r_{\rm eff}/a_0$ versus  $r_{\rm eff}$
is plotted
for  $\Omega^{++}_{ccc}\Omega^{++}_{ccc}$ ($^1S_0$) and  $\Omega_{sss}^-\Omega_{sss}^-$  ($^1S_0$) with (without) the Coulomb repulsion,
together with experimental values corresponding for $NN$($^3S_1-^3D_1$) and $NN$($^1S_0$).
In the case of   $\Omega^{-}_{sss}\Omega^{-}_{sss}$ ($^1S_0$), effects of the Coulomb repulsion to $a_0^{\rm C}$ and $r_{\rm eff}^{\rm C}$ are recalculated using eq.~(\ref{eq:Coulomb}) with the charge radius $r_d= 0.57$ fm for $\Omega_{sss}^-$~\cite{Can:2015exa}.
While all dibaryons and dibaryon candidates appear in unitary region,
 $\Omega^{++}_{ccc}\Omega^{++}_{ccc}$ ($^1S_0$) is the closest to unitarity among all, due to a subtle cancellation among the potential energy, the kinetic energy and the Coulomb repulsion. 
 
\subsection{5-4. $N\Omega_{sss}$ dibaryon}
Using the same gauge ensemble at the almost physical point, 
$N\Omega_{sss}^-$ interactions in the $^5S_2$ channel, which belongs to the {\bf 8} representation, have been extracted~\cite{HALQCD:2018qyu}.
At the almost physical point, $N\Omega^-_{sss}(^5S_2)$ can couple to octet-octet channels below the $N\Omega_{sss}^-$ threshold 
such as $\Lambda\Xi$ and $\Sigma\Xi$. Such couplings, however, are assumed to be small in this study,
since they couple with $D$-wave and are suppressed kinematically.

 \begin{figure}[h]
  \centering
  \includegraphics[width=0.49\textwidth]{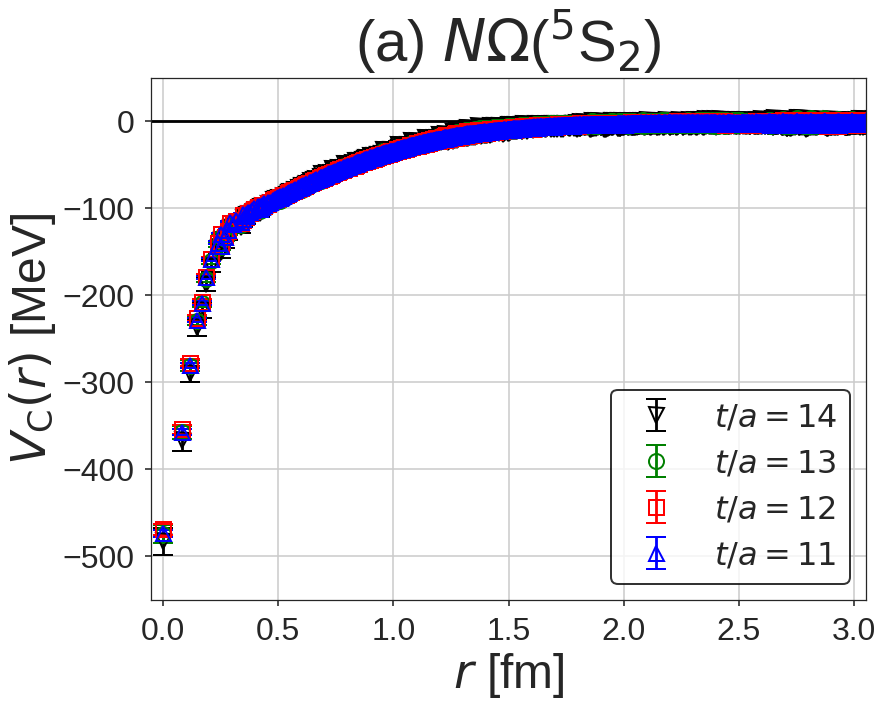}
  \includegraphics[width=0.49\textwidth]{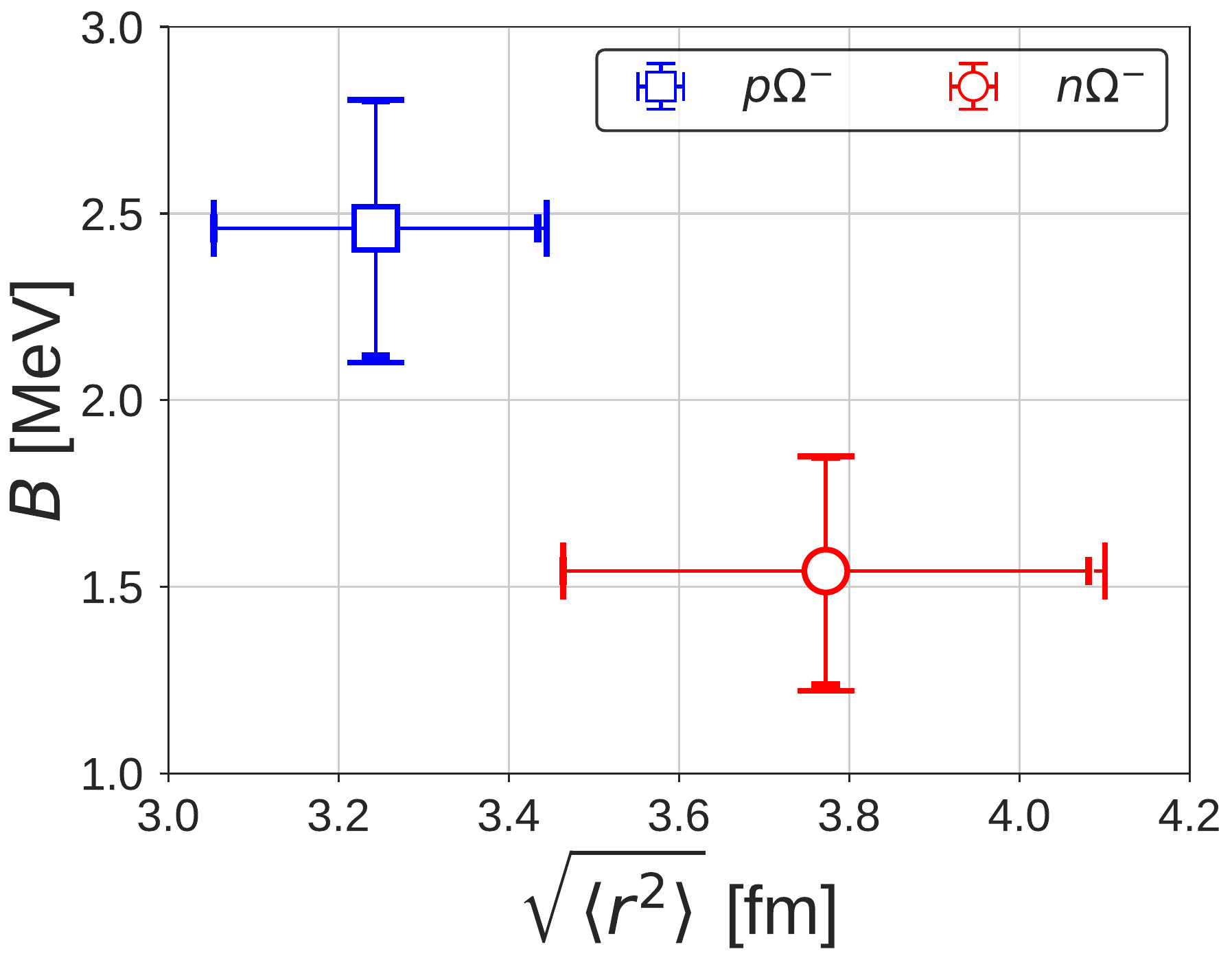}
  \caption{(Left)  The $N\Omega^{-}_{sss}$ potential in the $^5S_2$ channel at $t/a = 11$ (blue triangles), 12 (red squares), 13 (green circles), 14(black triangles) as a function of separation $r$ [fm], obtained in (2+1) lattice QCD simulations at  the almost physical point. 
    (Right)  The binding energy and the root-mean-square distance for the $n \Omega^{-}_{sss}$  with no Coulomb interaction (red open circle) and 
    $p^+\Omega^-$ with the Coulomb attraction (blue open square).
     Figures are  taken from~\cite{HALQCD:2018qyu}.
  }
\label{fig:NOmega} 
\end{figure}
Fig.~\ref{fig:NOmega} (Left) shows the $N\Omega_{sss}^-$ potential in the $^5S_2$ channel 
at $t/a = 11$ (blue triangles), 12 (red squares), 13 (green circle), 14 (black triangle),
which is attractive at all distances without repulsive core.
It is found that this potential produces one bound state in this channel,
as in the case of the flavor singlet potential in the flavor SU(3) limit. 
The binding energy is extracted as $B =1.5(3)$ MeV for $n \Omega_{sss}^-$ in the absence of  the Coulomb interaction or
 $B = 2.5(4)$ MeV for $p^+  \Omega_{sss}^-$  with the Coulomb attraction, 
 where  the Coulomb attraction is estimated by adding the Coulomb potential $ - \alpha_e/r$ between point-like proton and $\Omega_{sss}^-$. 
 See Fig.~\ref{fig:NOmega} (Right) for the  binding energy and the root-mean-square distance for $n\Omega_{sss}^-$ (red circle) and
 $p^+  \Omega_{sss}^-$ (blue square).
 These binding energies at the almost physical point are much smaller than $B = 19(5)^{+12}_{-2}$ MeV 
 at a heavier pion mass $m_\pi =875$ MeV~\cite{HALQCD:2014okw}.
Scattering parameters for $n \Omega_{sss}^-$  (no Coulomb) become
\beqa
a_0 =5.3 (5) {\rm fm}, \quad r_{\rm eff} =1.3(2) {\rm fm}, \quad {r_{\rm eff}\over a_0} \simeq 0.24,
\eeqa
showing that this system is also close to unitarity.

\section{\textit{6. Summary}}
In this chapter, recent progresses are given in lattice QCD studies on baryon-baryon interactions by the HAL QCD potential method. 
A key observation is that the NBS wave function of two baryons encodes information of the scattering phase shift between two baryons
below the corresponding inelastic threshold  in its asymptotic behavior outside the range of interactions.
Potentials at the interaction region can then be defined through the NBS wave function so as to reproduce the scattering phase shift by construction,
where non-locality of potentials is handled by the derivative expansion. 
Furthermore, potentials can be extracted from time dependences as well as space dependences of two-baryon correlation functions
without the ground state saturation, as potentials are constructed in the energy-independent way.

Various numerical results on baryon-baryon interactions in lattice QCD are introduced in this chapter.
They include 
$NN$ potentials in the parity-even sector such as $^1S_0$ and $^3S_1 - ^3D_1$ at heavy pion masses
as well as preliminary results of them  at the almost physical point,
$NN$ potentials in the parity-odd sector including the spin-orbit potential corresponding to NLO in the derivative expansion,
three-nucleon potential, baryon-baryon interaction and $H$-dibaryon in the flavor SU(3) symmetric limit, 
the fate of the $H$-dibaryon at the almost physical point, and
predictions on various dibaryons using the ensemble at the almost physical point.

Results in this chapter indicate a clear pathway from (lattice) QCD to nuclear physics.
While extractions of potentials at the physical point are still big challenges in lattice QCD,
continuous theoretical developments as well as progresses of supercomputers will further deepen the connection between lattice QCD and nuclear physics, whose outcomes are expected to play key roles to understand not only structures and dynamical properties of nuclei
but also nuclear astrophysical phenomena such as supernova explosions, mergers of binary neutron stars and 
nucleosynthesis associated with these explosive phenomena.    
   
\section*{Acknowledgement}
This work is supported in part by the Grand-in-Aid of the Japanese Ministry of Education, Sciences and Technology, Sports and Culture (MEXT)
for Scientific Research (Nos. JP18H05236, JP18H05407, JP19K03879, JP22H00129), 
by a priority issue (Elucidation of the fundamental laws and evolution of the universe) to be tackled by using Post ``K" Computer,
by ``Program for Promoting Researches on the Supercomputer Fugaku''
 (Simulation for basic science: from fundamental laws of particles to creation of nuclei),
and by Joint Institute for Computational Fundamental Science (JICFuS).
The authors thank members of the HAL QCD Collaboration
  for providing lattice QCD results
  and for fruitful collaborations based on which this paper is prepared.
    Figures~\ref{fig:NN_singlet} and \ref{fig:NN_triplet} are taken from~\cite{Ishii:2013ira},
    under the term of \cite{CC_BY-NC-ND_4.0}.
    Figure~\ref{fig:NN_K} is taken from~\cite{Doi:2017zov},
    under the term of~\cite{CC_BY_4.0}.
    Figures~\ref{fig:pot_odd} and \ref{fig:phase_odd} are taken from~\cite{Murano:2013xxa},
    under the term of~\cite{CC_BY_3.0}.
    Figures~\ref{fig:pot_SU3} and \ref{fig:pot_H} are reprinted from~\cite{Inoue:2011ai},
    and
    Figures~\ref{fig:phase_LL_NX} and \ref{fig:phase_NX} are reprinted from~\cite{HALQCD:2019wsz},
    with permission from Elsevier.
    Figure~\ref{fig:NOmega} is taken from~\cite{HALQCD:2018qyu},
    under the term of~\cite{CC_BY_4.0}.
%

%
%
%
%

\end{document}